\documentclass[%
 reprint,
superscriptaddress,
 amsmath,amssymb,
 aps,
 prx,
]{revtex4-2}

\usepackage{amsthm}

\usepackage{graphicx}
\usepackage{dcolumn}
\usepackage{bm}
\usepackage{physics}
\usepackage{booktabs}
\usepackage{enumerate}

\usepackage[pdftex,pagebackref=false]{hyperref} 
\hypersetup{
    colorlinks=true,        
    linkcolor=blue,         
    citecolor=red,        
    filecolor=magenta,      
    urlcolor=cyan           
}

\usepackage{tikz}
\usetikzlibrary{arrows.meta,bending,calc,fit}
\definecolor{secblue}{RGB}{28,75,180}
\definecolor{secfill}{RGB}{232,239,252}
\definecolor{gauss}{RGB}{245,160,0}
\definecolor{nongauss}{RGB}{225,40,30}
\newlength{\figwidth}
\newlength{\figheight}

\usepackage{xcolor}
\usepackage{cleveref}
\usepackage{placeins}
\usepackage{orcidlink}

\newcommand{\I}{\mathrm{i}}
\newcommand{\id}{\mathbb{I}}
\newcommand{\ad}{\operatorname{ad}}

\newtheorem{proposition}{Proposition}
\newtheorem{lemma}{Lemma}
\newtheorem{definition}{Definition}

\newcommand{\ethzurich}{Institute for Theoretical Physics, ETH Z\"urich, 8093 Z\"urich, Switzerland}
\newcommand{\qumami}{Quantum Mathematics, Algorithms and Machine Intelligence (QUMAMI), UPM, Spain}
\newcommand{\upm}{Universidad Politécnica de Madrid, Spain}

\newcommand{\orcidmatteo}{\orcidlink{0000-0002-9426-0377}}
\newcommand{\orcidjannes}{\orcidlink{0000-0001-7491-3660}}
\newcommand{\orcidjuan}{\orcidlink{0000-0001-7263-3462}}
\newcommand{\orcidadrian}{\orcidlink{0000-0001-5430-6468}}
\newcommand{\orcidarnav}{\orcidlink{0009-0000-1712-8797}}
\newcommand{\mycomment}[1]{}
\usepackage{multirow}

\begin{document}

\title{Dirac Propagation for Simulating Fermionic Observables under Particle-Number-Conserving Unitaries}
\author{G Arnav Naidu$^{\orcidarnav}$}
\thanks{These authors contributed equally to this work.}
\affiliation{\ethzurich}
\author{Matteo D'Anna$^{\orcidmatteo}$}
\thanks{These authors contributed equally to this work.}
\affiliation{\ethzurich}
\author{Adrián Pérez-Salinas$^{\orcidadrian}$}
\affiliation{\ethzurich}
\affiliation{\qumami}
\affiliation{\upm}
\author{Jannes Nys$^{\orcidjannes}$}
\affiliation{\ethzurich}
\author{Juan  Carrasquilla$^{\orcidjuan}$}
\affiliation{\ethzurich}

\begin{abstract}
Simulating the dynamics of interacting fermions remains a central computational challenge across physics and chemistry. We introduce Dirac propagation, a classical method that simulates local observables in fermionic systems subject to particle-number-conserving unitaries. We show that, in the operator expansion, strings with weights above the total particle number can be discarded without incurring any error in the final expectation value. This exact cutoff bounds the number of retained strings polynomially in the number of modes at fixed particle number, including for interacting dynamics. The formalism combines a compact representation with explicit expressions for interacting gates and exact treatment of number-conserving fermionic Gaussian and Clifford unitaries. This framework offers an approach to simulating fermionic many-body dynamics and classically benchmarking fermionic quantum circuits.
\end{abstract}
\maketitle

\section{Introduction}

Predicting the dynamics of interacting fermions is a central computational challenge across physics and chemistry~\cite{Keldysh:1964ud,aoki2014nonequilibrium,bauer2020quantum,bauer2023highenergy,abrams1997simulation,kivlichan2018quantum,rocca2024symmetry,lee2021hypercontraction}. Yet these dynamics provide valuable physical insight into thermalization in correlated materials~\cite{PolkovnikovEtAl2011,eckstein2009thermalization,rigol2008thermalization,moeckel2010crossover}, ultrafast charge redistribution in molecules~\cite{LiEtAl2020RealTime,runge1984density,bruner2017attosecond}, and optical control of magnetic order~\cite{KirilyukEtAl2010,beaurepaire1996ultrafast,peng2018dynamics,baiardi2021electron}, while also providing access to excitation spectra~\cite{li2020realtime,YabanaBertsch1996,nascimento2016linear}. Recent advances in analog Fermi-Hubbard simulators~\cite{Hirthe_2023,Bourgund_2025,XuEtAl2025Hubbard} and native gates for digital fermionic quantum computation~\cite{BojovicEtAl2026Gates} motivate the development of reliable classical simulation of fermionic matter as benchmarks for state preparation and dynamics.

These applications require measuring or predicting time-dependent observables such as densities, currents, spin correlations, and dipole moments. Heisenberg picture methods have recently emerged as an effective route to the dynamics of such local observables. Sparse Pauli dynamics, or Pauli propagation (PP)~\cite{Rall2019Simulation, begusic2024Fast, rudolph2025paulipropagationcomputationalframework, BegusicChan2025}, evolves the observable itself by expanding the Heisenberg picture operator in the Pauli basis. Originally developed to simulate many-qubit dynamics and quantum circuits \cite{begusic2024Fast,rudolph2023classicalsurrogatesimulationquantum,fontana_classical_2025}, the approach was extended to fermionic degrees of freedom through Majorana propagation (MP)~\cite{DAnnaNysCarrasquilla2026,MillerEtAl2025MajoranaPropagation}. Additionally, PP and MP were employed in variational settings to prepare the ground state of qubit and fermionic systems~\cite{chakraborty2026scalablequantumcircuitgeneration, Shrikhande2026, Lin2026,santona2026hardwareefficientgroundstatepreparationusing}. While interactions generate increasingly large expansions that make the methods numerically infeasible, operator growth is limited by truncating the expansion.

Here, we introduce Dirac propagation (DP), a formalism specifically designed for the simulation of particle-number-preserving dynamics acting on states with a fixed number of fermions. Such dynamics are ubiquitous in fermionic simulations, ranging from lattice models to molecular systems. Our approach expands observables in canonically ordered products of fermionic Dirac operators and establishes exact transformation rules based on recent theoretical progress~\cite{EvangelistaMagoulas2025,MagoulasEvangelista2026}.

We show that for simulating a system with $N$ fermions, for $N\in\mathcal O(1)$, the DP formalism gives exact results with a number of terms that scales polynomially in the number of fermionic modes $M$
\begin{equation}
  T_{N}=\sum_{q=0}^{\min(N,M-N)}\binom{M}{q}^{2}\in \tilde{\mathcal O}\left(M^{2\min(N,M-N)}\right),
\end{equation}
as opposed to the exponential scaling of general propagation algorithms \cite{rudolph2025paulipropagationcomputationalframework,DAnnaNysCarrasquilla2026}. To establish this more favorable scaling, we introduce the notion of Dirac-weight and show that, for initial states with $N$ particles, contributions with Dirac-weight greater than $N$ can be discarded without affecting the final expectation value. The scaling becomes exponential if $N$ is a constant filling fraction of $M$. 

Additionally, we show that number-preserving fermionic Gaussian rotations permit exact propagation of observables of bounded weight in this computational framework as well~\cite{DiVincenzoTerhal2005}. We use DP to simulate the dynamics of the $2$D Fermi-Hubbard model across interacting regimes, as well as to quench a molecular system with basis sets of increasing size. Together with our theoretical findings, these results show that DP provides reliable estimates for simulating the evolution of particle-preserving fermionic systems.

This work is organized as follows. Section \ref{sec.methodology} presents the formalism of Dirac propagation. The computational complexity of the formalism is investigated in  Section \ref{sec.complexity}. We apply Dirac propagation in Section \ref{sec.experiments}. Finally, we draw our conclusions in Section \ref{sec.conclusions}.

\section{Methodology}\label{sec.methodology}
We focus on computing expectation values of number-conserving local observables $O$ for an initial-state density operator $\rho$ with $N$ fermions and $M$ modes, evolved under number-conserving unitaries $U$. Under such restrictions, the  Fock space decomposes into fixed-particle sectors as
\begin{equation}
    \mathcal F = \bigoplus_{N=0}^M \mathcal F_N. 
\end{equation}

Following PP and MP, Dirac propagation represents and propagates observables by expanding them in an operator basis. Specifically, we use strings of Dirac operators, denoted by $\eta(v)$ and introduced in Section~\ref{sec:representation}, to express observables as
\[
O = \sum_v c_v \eta(v).
\]
Under Heisenberg evolution, we write the evolved observable as
\begin{align}
  O(t)&\equiv U^\dagger(t)OU(t)
       =\sum_v c_v(t)\eta(v),
       \label{eq:evolved-observable}
\end{align}
where the coefficients $c_v(t)\in\mathbb C$. 
The time-dependent expectation values $\langle O \rangle_t$ for an initial state $\rho$ with $N$ fermions, are given by
\begin{align}
\langle O\rangle_t
      &\equiv\Tr_N\!\left[\rho O(t)\right]
       =\sum_v c_v(t)
        \Tr_N\!\left[\rho\eta(v)\right],
       \label{eq:evolution-op-expansion}
\end{align}
where $\Tr_N$ denotes the trace restricted to the $N$-particle subsector $\mathcal{F}_N$ of the Fock space. 

To ensure computational tractability, we restrict the method to observables $O$ with a sparse representation, i.e., at most $\mathcal O(\operatorname{poly}(M))$ non-vanishing terms. Examples of such sparse operators are local observables.

\subsection{Representation}\label{sec:representation}
We now introduce the Dirac basis using fermionic Dirac operators. Consider the Fock space $\mathcal F$ of $M$ ordered fermionic modes with a fixed mode order $1<\cdots<M$. The fermionic creation and annihilation operators $f^\dagger_k, f_k$ (also called Dirac operators~\cite{Dirac:1927dy}) obey
\begin{equation}
  \{f_j,f_k^\dagger\}=\delta_{jk}\id,
  \qquad
  \{f_j,f_k\}=\{f_j^\dagger,f_k^\dagger\}=0,
  \label{eq:car}
\end{equation}
where the subscripts indicate the corresponding fermionic modes. The number operator is $n_k=f_k^\dagger f_k$ at each mode and the total number operator is $\hat{N} = \sum_k n_k$. Particle-conserving unitaries $U$ fulfill the condition $[U,\hat{N}]=0$.

\subsubsection{Canonical form}
We introduce a canonical form for Dirac operators so each operator has a unique representation. We start by discussing the case where we have a single fermionic mode. The set $\{0,\mathbb I, f,f^\dagger,f f^\dagger, f^\dagger f\}$ has the following multiplication table (row $\times$ column)
\begin{equation}\label{eq:multiplication_table}
\begin{array}{c|cccccc}
 & 0 & \mathbb I & f & f^\dagger & ff^\dagger & f^\dagger f\\
 \hline 
 0 & 0 & 0&0 & 0&0 & 0  \\
 \mathbb I & 0 & \mathbb I & f & f^\dagger & f f^\dagger & f^\dagger f\\
 f &0 & f&0 &ff^\dagger &0 & f  \\
 f^\dagger & 0& f^\dagger & f^\dagger f &0 &f^\dagger & 0  \\
 f f^\dagger & 0& f f^\dagger & f &0 &f f^\dagger & 0  \\
 f^\dagger f & 0&  f^\dagger f & 0 & f^\dagger & 0 & f^\dagger f \\

\end{array}
\end{equation}
For example, $f \times f f^\dagger = f^2 f^\dagger = 0$.
This structure forms a monoid. The main difference between a monoid and a group is that, in a monoid, not all elements have an inverse in the set. We select a set of elements of the monoid to build a (non-orthogonal) basis of the single-mode fermionic system 
\begin{align}
  \mathcal D_1=\{\id,f,f^\dagger,n=f^\dagger f\}.
  \label{eq:local-basis}
\end{align}
We refer to $\mathcal D_1$ as the local canonical Dirac operator basis. This basis $\mathcal D_1$ imposes a canonical ordering on the operators, in which we express $f f^\dagger $ as 
\begin{equation}
f f^\dagger = \mathbb I - n.
\end{equation}

For a system with $M$ fermionic modes, we introduce the Dirac-operator string
\begin{equation}
  \eta(v)=(f_1^\dagger)^{v_1}(f_1)^{v_2} \cdots (f_M^\dagger)^{v_{2M-1}}(f_{M})^{v_{2M}},
  \label{eq:eta}
\end{equation}
indexed by the $2M$-bitstring $v \in \{0, 1\}^{2M}$. This allows us to describe the \textit{canonical operator basis} 
\begin{equation}
    \mathcal D_M = \{ \eta(v): v \in \{0,1\}^{2M}\}.
\end{equation}
Each string also corresponds to a monomial (a single ordered product of creation and annihilation operators), whereas we will use the term
\emph{Dirac expansion} for a linear combination of such elements. Since we focus on the case of number-preserving observables and unitaries, a Dirac-operator string will have equally many creation and annihilation operators.
A basis for such a particle-preserving operator space is given as
\begin{equation}
    \mathcal B_M = \left\{ \eta(v) \in \mathcal D_M : \sum_{k=1}^Mv_{2k-1}=\sum_{k=1}^Mv_{2k}\right\}.
\end{equation}
Strings in $\mathcal D_M,\mathcal B_M$ are non-orthogonal; see Appendix \ref{app:normal-ordering} for more details.

\subsubsection{Dirac-weight}

Distinct from Pauli or Majorana weights~\cite{rudolph2025paulipropagationcomputationalframework, DAnnaNysCarrasquilla2026}, we define a new weight for Dirac strings and expansions to explain the improved scaling behavior of our method.

\begin{definition}\label{def:weight-string}
We define the \emph{Dirac-weight} $w(\eta(v))$ of a string $\eta(v)\in \mathcal B_M$ as the number of annihilation operators $f_k$ within the string
\begin{align}
    w(\eta(v)) \equiv w(v) = \sum_{k=1}^M v_{2k} \label{eq:weight}.
\end{align}
\end{definition}
\noindent We will also refer to the Dirac-weight as simply \emph{weight}. 
\begin{definition}\label{def:weight-expansion}
For a number-conserving Dirac expansion $G = \sum_v c_v \eta(v)$ we refer to its Dirac-weight as 
\begin{align}
    w(G) &:= \min_{v | c_v \neq 0} w(\eta(v)).
\end{align}
\end{definition}

For the case of $s$ fermionic species (e.g. with different projections), we write the Fock space as
\begin{equation}
   \mathcal F=\bigoplus_{N_1=0}^{M_1}\cdots \bigoplus_{ N_s=0}^{M_s}  \mathcal F_{N_1,\cdots, N_s},
\end{equation}
where $\mathcal F_{N_1,\cdots, N_s}$ is the sector with $N_j$ particles of species $j$, and $M_j$ is the number of modes of species $j$. Dirac strings \eqref{eq:eta} are built by choosing an arbitrary ordering of the $M=M_1+\cdots+M_s$ modes.
Let $S_j\subset [M]$ be the index set for the modes of species $j$. We then define the per-species weight $w_j(v)$ of a Dirac string $\eta(v)$ as the number of annihilation operators $f_k$ on modes $k$ associated with the species $j$,
\begin{equation}
    w_j(v) = \sum_{k \in S_j} v_{2k}.
\end{equation}
We will refer to this per-species weight as $j$-weight.

\subsubsection{Overlap with Fock states}
After propagating the observable, evaluating \eqref{eq:evolution-op-expansion} requires the overlaps between Dirac strings and the initial state. Here we focus on the common case of Fock states. Consider a Dirac string $\eta(v)\in\mathcal B_M$, and let $I=\{k: v_{2k}=1\}$ and $J=\{k: v_{2k-1}=1\}$ be index sets of its annihilation and creation modes, respectively. Let $K=\{k_1,\cdots, k_p\}$ and $L=\{l_1,\cdots, l_{q}\}$ be index sets that store occupied modes from two Fock states with $(k_1 < \ldots < k_p)$ and $(l_1 < \ldots < l_q)$
\begin{equation}\label{eq:fock-states}
\begin{split}
    \ket K &= f_{k_1}^\dagger \cdots f_{k_{p}}^\dagger\ket 0\\
    \ket L &= f_{l_1}^\dagger \cdots f_{l_{q}}^\dagger\ket 0.
    \end{split}
\end{equation}
The expression for matrix elements is given as
\begin{equation}\label{eq:overlap}
    |\bra{K} \eta(v) \ket{L}|=\begin{cases}
        1 & \textrm{if } I\subseteq L \textrm{ and } J\subseteq K \textrm{ and } L\backslash I = K \backslash J\\
        0 & \textrm{otherwise}
    \end{cases}
\end{equation}
From the definition of weight, it is natural to see that any string with weight $w(\eta(v))>|L|$ would give $I\nsubseteq L$. Also, since $\eta(v)\in\mathcal B_M$, $|I|=|J|=w(\eta(v))$, and we find
\begin{equation}
    \bra{K} \eta(v) \ket{L} = 0 \quad \textrm{if } w(\eta(v))> \min(|K|, |L|).
\end{equation}
Thus, if $\rho$ is an initial state with $N$ particles,
\begin{equation}\label{eq:trace_rho}
    \mathrm{Tr}(\eta(v)\rho)= 0 \quad \textrm{if }w(\eta(v))>N.
\end{equation}
In other words, all terms in the Dirac expansion with weight higher than $N$ do not contribute to an expectation value \eqref{eq:evolution-op-expansion}, as we will discuss in more detail in Section~\ref{sec:weight_monot}.

\subsection{Propagation rules}
Having introduced the Dirac basis, we now return to the task \eqref{eq:evolved-observable} of propagating observables subject to a unitary $U$ in the Heisenberg picture. 
In terms of a Dirac basis expansion, we use the expression of the exponential of the adjoint action (with $\operatorname{ad}_X(Y) = [X,Y]$)~\cite{rycroft_theoretical_physics_2}
\begin{equation}\label{eq:hadamard-series}
\begin{split}
    e^X Y e^{-X} &= e^{\operatorname{ad}_X}Y \\
    &= \sum_{n=0}^{\infty} \frac{1}{n!}
      \operatorname{ad}_X^n(Y) \\
    \end{split}
\end{equation}
To propagate a canonical Dirac string $\eta(v)\mapsto U^\dagger\eta(v)U$, we consider a Hermitian generator constructed from a single Dirac string $\eta(g)$
\begin{align}\label{eq:string-generator}
G_+(g)&\equiv\eta(g)+\eta(g)^\dagger \\
G_-(g)&\equiv-\I(\eta(g)-\eta(g)^\dagger),
\end{align}
with $\eta(g)\in\mathcal B_M$ and we introduce \[U_\pm\equiv e^{-\I\theta G_\pm(g)}.\]
For a canonical string $\eta(v)$, the nested commutators satisfy~\cite{EvangelistaMagoulas2025}
\begin{equation}
    \operatorname{ad}_{G_\pm}^3\big(\eta(v)\big)
    =\kappa\,\operatorname{ad}_{G_\pm}\big(\eta(v)\big),
    \qquad \kappa\in\{1,4\},
    \label{eq:commutator-closure}
\end{equation}
where the appropriate value of $\kappa$ depends on the generator and the propagated string; see Appendix~\ref{app:expressions}.
This closure allows \eqref{eq:hadamard-series} to be resummed exactly, yielding the \emph{Dirac propagation splitting rule}~\cite{EvangelistaMagoulas2025,MagoulasEvangelista2026}
\begin{equation}
\begin{split}
    U^\dagger_\pm\eta(v)U_\pm
    =\;&\eta(v)
    +\I\frac{\sin(\sqrt{\kappa}\theta)}{\sqrt{\kappa}}
      [G_\pm,\eta(v)]\\
    &+\frac{\cos(\sqrt{\kappa}\theta)-1}{\kappa}
      [G_\pm,[G_\pm,\eta(v)]].
\end{split}
\label{eq:splitting-rule}
\end{equation}
A string that commutes with the generator is unchanged. Otherwise, its evolution is determined entirely by the first and second nested commutators, with the trigonometric coefficients accounting for all higher orders. The evolved operator therefore lies in the span of $\eta(v)$, $[G_\pm,\eta(v)]$, and $[G_\pm,[G_\pm,\eta(v)]]$. Notice that each commutator may contain several Dirac strings. Also note that by expanding the closed form \eqref{eq:splitting-rule} in powers of $\theta$ with $\sin \theta = \theta + \order{\theta^3}$ and $1-\cos \theta = \tfrac{\theta^2}{2} + \order{\theta^4}$, we see that a canonical Dirac string is split into itself and additional terms of order $\theta$ or higher.

\paragraph{Example.} As a concrete example of the DP splitting rule, we consider the rotation under a density-density interaction $U_{nn} = \exp(-\I\theta n_in_j)$ applied to a creation operator. We obtain
\begin{equation}\label{eq:splitting_example}
  U_{nn}^{\dagger}f_k^\dagger U_{nn}
  =
  \begin{cases}
    f_i^\dagger+
    \left(e^{\I\theta}-1\right)f_i^\dagger n_j,
      & k=i,\\
    f_j^\dagger+
    \left(e^{\I\theta}-1\right)f_j^\dagger n_i,
      & k=j,\\
    f_k^\dagger,
      & k\notin\{i,j\},
  \end{cases}
\end{equation}
since the commutators are given as
\begin{align}\label{eq:splitting_example_commutators}
    [n_i n_j,f_i^\dagger] = f_i^\dagger f_i f_j^\dagger f_j f_i^\dagger-\underbrace{f_i^\dagger f_i^\dagger f_i f_j^\dagger f_j}_{=0} = f^\dagger_i n_j \\
    [n_i n_j,[n_i n_j,f_i^\dagger]] = [n_i n_j, f_i^\dagger n_j] = f_i^\dagger n_j
\end{align}
and hence \eqref{eq:splitting-rule} ($\kappa=1$) yields the stated expression \eqref{eq:splitting_example}. Notice that it is sufficient to consider this case of evolving a single creation operator, as the action of $U$ on general Dirac strings can be decomposed into the action on the single creation and annihilation operators
\begin{equation}
  U^\dagger\eta(v)U
  =
  \prod_{k=1}^{M}
  \left(U^\dagger f_k^\dagger U\right)^{v_{2k-1}}
  \left(U^\dagger f_k U\right)^{v_{2k}}.
\end{equation}
Appendix~\ref {app:expressions} gives explicit expressions for the splitting rule of other typical fermionic interactions.

\paragraph{Comparison with the Majorana basis.} We compare the Dirac splitting rule in Eq.~\eqref{eq:splitting-rule} to the Majorana propagation splitting rule~\cite{DAnnaNysCarrasquilla2026}.
The latter follows from the same resummation: for two anticommuting Hermitian Majorana strings $\mu(v)$ and $\mu(\tilde v)$, the generator $G = \mu(\tilde v)/2$ satisfies $G^2 = 1/4$ and $\{G,\mu(v)\} = 0$, so that $[G,[G,\mu(v)]] = 4G^2\mu(v) = \mu(v)$. Hence $\kappa = 1$, the double commutator collapses onto the original string, and
\begin{equation}\label{eq:MPsplit}
    e^{\I\theta \mu(\tilde v)/2}\, \mu(v)\, e^{-\I\theta \mu(\tilde v)/2} = \cos(\theta)\,\mu(v) + \I\sin(\theta)\,\mu(\tilde v)\mu(v),
\end{equation}
where $\I\mu(\tilde v)\mu(v) = \pm\mu(v+\tilde v)$ is again a Majorana string. A non-commuting Majorana rotation therefore splits one string into two. In Dirac propagation, the generator does not square to a multiple of the identity. Whenever $\eta(g)$ contains an unpaired creation or annihilation operator, $\eta(g)^2 = 0$ and $G_\pm^2 = \eta(g)\eta(g)^\dagger + \eta(g)^\dagger\eta(g)$ is a sum of projectors. As a result, in contrast to the MP case, a Dirac string $\eta(v)$ is in general not an eigenoperator of $\ad_{G_\pm}^2$. For the hopping generator $G = f_1^\dagger f_2 + f_2^\dagger f_1$ and $\eta(v) = n_1$, for instance, one finds $[G,[G,n_1]] = 2(n_1 - n_2)$. The double commutator thus generates strings distinct from $\eta(v)$, and all three terms in \eqref{eq:splitting-rule} must be retained.

\section{Complexity}\label{sec.complexity}
\subsection{Weight monotonicity}\label{sec:weight_monot}
The weight $w(v)$ \eqref{eq:weight} is central to reducing the complexity of simulations of expectation values \eqref{eq:evolution-op-expansion}. For a Dirac expansion, this is the smallest particle sector on which the Dirac expansion acts nontrivially, as we prove in Appendix \ref{sec:proof_weight_invariance}.

Additionally, the Dirac basis has the following crucial property: if we consider a number-preserving unitary evolution acting on a Dirac expansion $G$ with $w(G)=k$, the weight is invariant, i.e.
\begin{align}
    w(U^\dagger G U)= w(G) = k. \label{eq:weight-invariant}
\end{align}
While \eqref{eq:weight-invariant} guarantees that there will be at least one string in $U^\dagger G U$ with weight $k$, it will in general contain strings of weight higher than $k$; see Figure \ref{fig:weight_sectors}. We prove this statement in Appendix \ref{sec:proof_weight_invariance}.
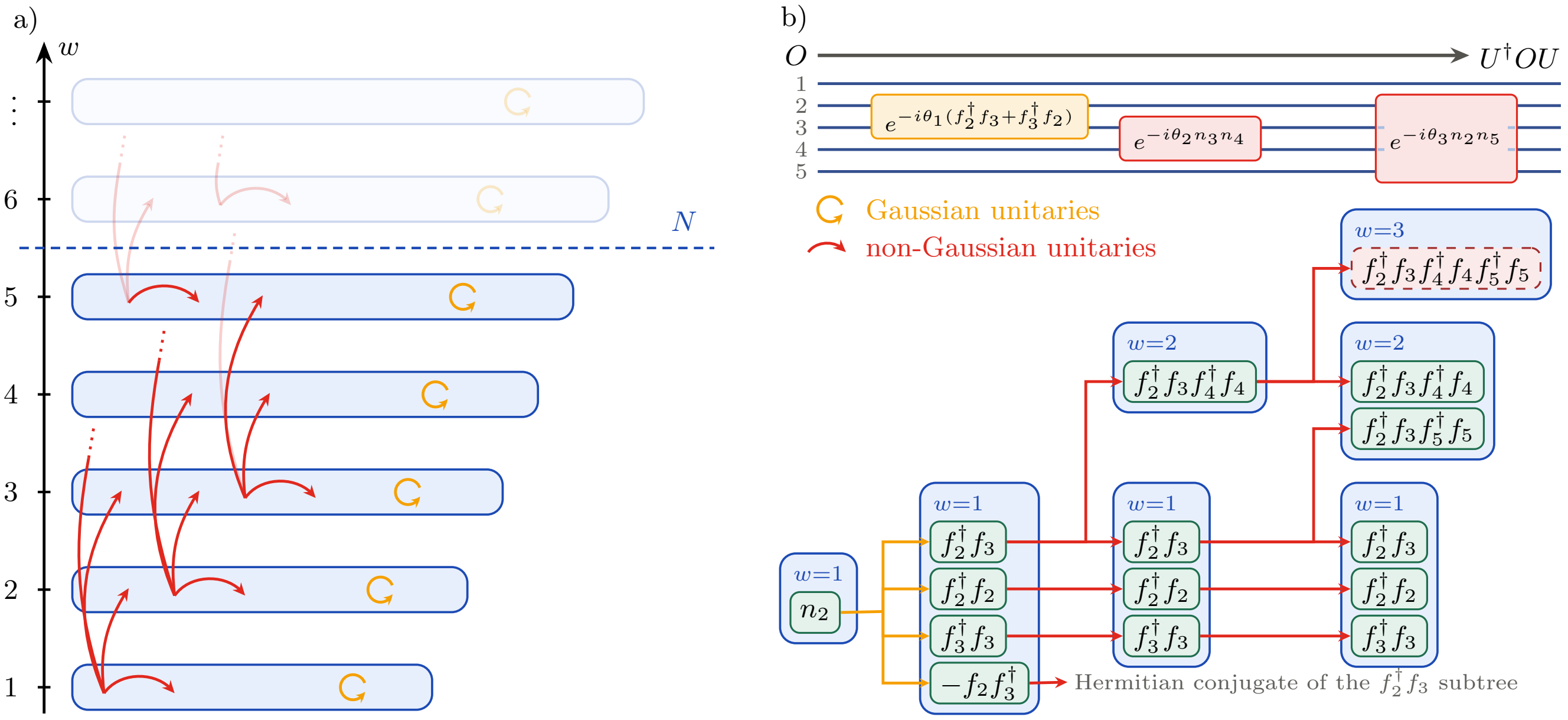
\begin{figure*}
\centering
\setlength{\figheight}{0.42\textwidth}
\begin{minipage}[b]{0.49\textwidth}
  \centering
  \setlength{\figwidth}{\linewidth}



\ifdefined\figwidth\else
  \newlength{\figwidth}\setlength{\figwidth}{10.5cm}
\fi
\ifdefined\figheight\else
  \newlength{\figheight}\setlength{\figheight}{9.5cm}
\fi

\begin{tikzpicture}[font=\small,
    x=0.09\figwidth, y=0.1117\figheight,
    >=Stealth,
    sector/.style={draw=secblue, fill=secfill, line width=0.7pt, rounded corners=5pt},
    gloop/.style={gauss, line width=0.9pt, -{Stealth[length=3.2pt, width=3pt, bend]}},
    ng/.style={nongauss, line width=0.9pt, -{Stealth[length=4pt, width=3.6pt, flex]}},
    ngmore/.style={nongauss, line width=0.9pt, dotted},
    fade/.style={opacity=#1},
  ]
 
  \def\lr{4pt}          
  \def\dy{1.3}         
  \def\hgt{0.6}        
  \def\xo{0.40}        
  \def\base{4.6}       
  \def\grow{0.5}       
  \def\nmax{7}         
  \def\mthr{5}         
  \def\fadeop{0.22}    
 
  \newcommand{\selfloop}[1]{\draw[gloop] ([shift={(20:\lr)}] #1) arc (20:340:\lr);}
 
  \draw[->, line width=0.9pt] (0,{\dy-\hgt/2-0.05}) -- (0,{\nmax*\dy+0.8}) node[anchor=north west, xshift=4pt, inner sep=0pt] {$w$};
  \node[anchor=south west, inner sep=0pt, yshift=2pt] at (-10pt,{\nmax*\dy+0.8}) {a)};
  \foreach \w in {1,...,\nmax} {
    \draw[line width=0.8pt] (-0.09,{\w*\dy}) -- (0.09,{\w*\dy});
  }
  \foreach \w in {1,...,6} {
    \node[left=5pt] at (0,{\w*\dy}) {$\w$};
  }
  \node[left=5pt] at (0,{\nmax*\dy}) {$\vdots$};
 
  \foreach \w in {1,...,\nmax} {
    \pgfmathsetmacro{\wd}{\base+\grow*\w}
    \pgfmathsetmacro{\yc}{\w*\dy}
    \pgfmathsetmacro{\fade}{\w>\mthr ? \fadeop : 1}
    \begin{scope}[fade=\fade]
      \filldraw[sector] (\xo,{\yc-\hgt/2}) rectangle ({\xo+\wd},{\yc+\hgt/2});
      \selfloop{{\xo+0.78*\wd},\yc}
    \end{scope}
  }
 
  \foreach \w/\sx/\nup in {1/0.85/2, 2/1.85/2, 3/2.85/2, 5/1.2/1, 6/2.5/0} {
    \pgfmathsetmacro{\sy}{\w*\dy-0.08}        
    \pgfmathsetmacro{\fadesrc}{\w>\mthr ? \fadeop : 1}
 
    \begin{scope}[fade=\fadesrc]
      \draw[ng] (\sx,\sy) to[bend left=55] ({\sx+1.0},\sy);
    \end{scope}
 
    \ifnum\nup>0
      \foreach \k in {1,...,\nup} {
        \pgfmathtruncatemacro{\wt}{\w+\k}
        \pgfmathsetmacro{\fadet}{\wt>\mthr ? \fadeop : 1}
        \pgfmathsetmacro{\ty}{\wt*\dy+0.02}
        \pgfmathsetmacro{\tx}{\sx+0.45-0.1*\k}
        \pgfmathsetmacro{\bendang}{12+8*\k}
        \begin{scope}[fade=\fadet]
          \draw[ng] (\sx,\sy) to[bend left=\bendang] (\tx,\ty);
        \end{scope}
      }
    \fi
 
    \def\stubang{80}      
    \def\stubbend{14}     
    \pgfmathtruncatemacro{\wn}{\w+\nup+1}
    \pgfmathsetmacro{\faden}{\wn>\mthr ? \fadeop : 1}
    \pgfmathsetmacro{\ey}{(\wn-1)*\dy+\hgt/2+0.15}
    \pgfmathsetmacro{\ex}{\sx+(\ey-\sy)*cot(\stubang+\stubbend)}
    \begin{scope}[fade=\faden]
      \draw[nongauss, line width=0.9pt] (\sx,\sy) to[bend left=\stubbend] (\ex,\ey);
      \draw[ngmore] (\ex,\ey) -- ++(\stubang:0.45);
    \end{scope}
  }
 
  \pgfmathsetmacro{\ym}{(\mthr+0.5)*\dy}
  \draw[secblue, densely dashed, line width=0.8pt] (-0.35,\ym) -- (9.5,\ym)
    node[above left=2pt, pos=1] {$N$};

\end{tikzpicture}%
%
\end{minipage}\hfill
\begin{minipage}[b]{0.49\textwidth}
  \centering
  \setlength{\figwidth}{\linewidth}

\ifdefined\figwidth\else
  \newlength{\figwidth}\setlength{\figwidth}{13cm}
\fi
\ifdefined\figheight\else
  \newlength{\figheight}\setlength{\figheight}{7.5cm}
\fi

\definecolor{wire}{HTML}{34508F}     
\definecolor{wirelight}{HTML}{A9BEE0}
\definecolor{keepfill}{HTML}{E4F2EB}
\definecolor{keepdraw}{HTML}{1F6E4E}
\definecolor{cutfill}{HTML}{FBEAEA}
\definecolor{cutdraw}{HTML}{A12E2E}
\definecolor{axis}{HTML}{55534E}
\definecolor{note}{HTML}{5F5E5A}
\providecolor{secblue}{RGB}{28,75,180}
\providecolor{secfill}{RGB}{232,239,252}
\providecolor{gauss}{RGB}{245,160,0}
\providecolor{nongauss}{RGB}{225,40,30}

\tikzset{
  gloop/.style    = {gauss, line width=0.9pt, -{Stealth[length=3.2pt, width=3pt, bend]}},
  ng/.style       = {nongauss, line width=0.9pt, -{Stealth[length=4pt, width=3.6pt, flex]}},
  bus/.style      = {gauss, line width=0.9pt},
  wireline/.style = {wire, line width=0.9pt},
  passwire/.style = {wirelight, line width=0.9pt},
  gatebox/.style  = {line width=0.6pt, rounded corners=2pt, inner sep=0pt},
  gatelabel/.style= {inner sep=1.5pt, font=\scriptsize},
  ggate/.style    = {draw=gauss, fill=gauss!15},
  glabel/.style   = {fill=gauss!15},
  nggate/.style   = {draw=nongauss, fill=nongauss!12},
  nglabel/.style  = {fill=nongauss!12},
  card/.style     = {line width=0.6pt, rounded corners=3pt,
                     minimum height=13pt, inner xsep=3pt, inner ysep=0pt},
  retained/.style = {card, draw=keepdraw, fill=keepfill},
  cut/.style      = {card, draw=cutdraw, fill=cutfill, dashed},
  notetext/.style = {text=note, font=\scriptsize, inner sep=2pt},
  sector/.style   = {draw=secblue, fill=secfill, line width=0.7pt, rounded corners=5pt, inner sep=3pt},
  wlabel/.style   = {text=secblue, font=\scriptsize, inner sep=1pt, anchor=north west},
}

\def\card#1#2#3#4{\node[#1, anchor=north west] (#2) at #3 {$#4$};}

\pgfdeclarelayer{bg}
\pgfsetlayers{bg,main}

\begin{tikzpicture}[x=1pt, y=1pt, font=\small]

\pgfmathsetlengthmacro{\xr}{\figwidth-1pt}                 
\pgfmathsetlengthmacro{\gx}{(\xr-178pt)/3}   
\pgfmathsetlengthmacro{\gy}{(\figheight-190pt)/3}       
\pgfmathsetlengthmacro{\xa}{24pt+\gx}                   
\pgfmathsetlengthmacro{\xb}{\xa+38pt+\gx}               
\pgfmathsetlengthmacro{\xc}{\xb+49pt+\gx}               
\pgfmathsetlengthmacro{\yax}{\figheight-5pt}            
\pgfmathsetlengthmacro{\ytc}{\figheight-46pt-\gy}       
\pgfmathsetlengthmacro{\ytb}{\ytc-28pt-\gy}             
\pgfmathsetlengthmacro{\yta}{\ytb-43pt-\gy}             
\newcommand{\ywire}[1]{\yax-9pt-7pt*(#1-1)}             

\node[anchor=south west, inner sep=0pt, yshift=2pt] at (0,\figheight) {b)};

\node[anchor=east, inner sep=0pt] (uou) at (\xr,\yax) {$U^\dagger O U$};
\draw[axis, line width=1.1pt, -{Stealth[length=6pt,width=5pt]}] (12,\yax) -- ([xshift=-3pt] uou.west);
\node[anchor=east, inner sep=1pt] at (10,\yax) {$O$};
\foreach \j in {1,...,5}{
  \draw[wireline] (12,{\ywire{\j}}) -- (\xr,{\ywire{\j}});
  \node[anchor=east, text=note, font=\scriptsize, inner sep=1pt] at (10,{\ywire{\j}}) {$\j$};
}
\node[gatebox, ggate, minimum width=69pt, minimum height=14pt] (g1) at ({\xa+16pt},{(\ywire{2}+\ywire{3})/2}) {};
\node[gatelabel, glabel] at (g1) {$e^{-i\theta_1(f^\dagger_2 f_3 + f^\dagger_3 f_2)}$};
\node[gatebox, nggate, minimum width=45pt, minimum height=14pt] (g2) at ({\xb+21.5pt},{(\ywire{3}+\ywire{4})/2}) {};
\node[gatelabel, nglabel] at (g2) {$e^{-i\theta_2 n_3 n_4}$};
\node[gatebox, nggate, minimum width=45pt, minimum height=28pt] (g3) at ({\xc+30.5pt},{(\ywire{2}+\ywire{5})/2}) {};
\foreach \j in {3,4} \draw[passwire] ({\xc+30.5pt-21.9pt},{\ywire{\j}}) -- ({\xc+30.5pt+21.9pt},{\ywire{\j}});
\node[gatelabel, nglabel] at (g3) {$e^{-i\theta_3 n_2 n_5}$};

\begin{scope}[shift={(16pt,{\figheight-54pt})}]
  \draw[gloop] ([shift={(20:4pt)}] 0,0) arc (20:340:4pt);
  \node[right=8pt, text=gauss] at (0,0) {Gaussian unitaries};
  \draw[ng] (-7pt,-13.7pt) to[bend left=55] (5pt,-13.7pt);
  \node[right=8pt, text=nongauss] at (0,-12pt) {non-Gaussian unitaries};
\end{scope}

\node[wlabel] (lr) at (3pt,{\yta-25.5pt}) {$w{=}1$};
\card{retained}{root}{(3pt,{\yta-34.5pt})}{n_2}

\node[wlabel] (la1) at (\xa,{\yta-3pt}) {$w{=}1$};
\card{retained}{a1}{(\xa,{\yta-12pt})}{f^\dagger_2 f_3}
\card{retained}{a2}{(\xa,{\yta-27pt})}{f^\dagger_2 f_2}
\card{retained}{a3}{(\xa,{\yta-42pt})}{f^\dagger_3 f_3}
\card{retained}{a4}{(\xa,{\yta-57pt})}{-f_2 f^\dagger_3}
\node[wlabel] (lb1) at (\xb,{\yta-3pt}) {$w{=}1$};
\card{retained}{b1}{(\xb,{\yta-12pt})}{f^\dagger_2 f_3}
\card{retained}{b2}{(\xb,{\yta-27pt})}{f^\dagger_2 f_2}
\card{retained}{b3}{(\xb,{\yta-42pt})}{f^\dagger_3 f_3}
\node[wlabel] (lb2) at (\xb,{\ytb-3pt}) {$w{=}2$};
\card{retained}{b5}{(\xb,{\ytb-12pt})}{f^\dagger_2 f_3 f^\dagger_4 f_4}
\node[wlabel] (lc1) at (\xc,{\yta-3pt}) {$w{=}1$};
\card{retained}{c1}{(\xc,{\yta-12pt})}{f^\dagger_2 f_3}
\card{retained}{c2}{(\xc,{\yta-27pt})}{f^\dagger_2 f_2}
\card{retained}{c3}{(\xc,{\yta-42pt})}{f^\dagger_3 f_3}
\node[wlabel] (lc2) at (\xc,{\ytb-3pt}) {$w{=}2$};
\card{retained}{c4}{(\xc,{\ytb-12pt})}{f^\dagger_2 f_3 f^\dagger_4 f_4}
\card{retained}{c5}{(\xc,{\ytb-27pt})}{f^\dagger_2 f_3 f^\dagger_5 f_5}
\node[wlabel] (lc3) at (\xc,{\ytc-3pt}) {$w{=}3$};
\card{cut}{c6}{(\xc,{\ytc-12pt})}{f^\dagger_2 f_3 f^\dagger_4 f_4 f^\dagger_5 f_5}

\begin{pgfonlayer}{bg}
  \node[sector, fit=(lr)(root)] {};
  \node[sector, fit=(la1)(a1)(a4)] {};
  \node[sector, fit=(lb1)(b1)(b3)] {};
  \node[sector, fit=(lb2)(b5)] {};
  \node[sector, fit=(lc1)(c1)(c3)] {};
  \node[sector, fit=(lc2)(c4)(c5)] {};
  \node[sector, fit=(lc3)(c6)] {};
\end{pgfonlayer}

\pgfmathsetlengthmacro{\xbus}{\xa-3pt-\gx/2}
\draw[bus] (root.east) -- (\xbus,{\yta-41pt});
\foreach \n in {a1,a2,a3,a4} \draw[gloop] (\xbus,{\yta-41pt}) |- (\n.west);
\foreach \s/\t in {a1/b1, a2/b2, a3/b3} \draw[ng] (\s.east) -- (\t.west);
\node[notetext, anchor=west] (hc) at ({\xa+44pt},{\yta-63.5pt})
  {Hermitian conjugate of the $f^\dagger_2 f_3$ subtree};
\draw[ng] (a4.east) -- (hc.west);
\draw[ng] (a1.east) -| ({\xa+38pt+\gx/2},0 |- b5.west) -- (b5.west);
\foreach \s/\t in {b1/c1, b2/c2, b3/c3, b5/c4} \draw[ng] (\s.east) -- (\t.west);
\draw[ng] (b1.east) -| ({\xb+49pt+\gx/2},0 |- c5.west) -- (c5.west);
\draw[ng] (b5.east) -| ({\xb+49pt+\gx/2},0 |- c6.west) -- (c6.west);

\end{tikzpicture}%
\end{minipage}
\caption{(a) Schematic representation of the absence of backflow in DP. For Gaussian unitaries, the weight of the strings in the Dirac expansion is fixed. In the general non-Gaussian regime, the weight of the strings in a Dirac expansion is non-decreasing, but the dynamics causes strings to increase their weight. Overlaps with $N$-particles initial states impose that strings $\eta(v)$ with $w(v)>N$ can be discarded without incurring any approximation error in the expectation value.
(b) Illustration of the absence of backflow and weight monotonicity in DP. An initial string $n_2=f^\dagger_2 f_2$ propagates through Gaussian and four-body gates. Dirac strings are written in the canonical ordering of Section~\ref{sec:representation}. The weight truncation threshold is set to $w_{\max}=2$, so the weight-$3$ string generated by the last gate is discarded. After the first hopping gate, the strings $f_2^\dagger f_3$ and $-f_2 f_3^\dagger = f_3^\dagger f_2$ are Hermitian conjugates of each other. Since Heisenberg evolution is an adjoint preserving map, their propagation trees are conjugate as well, and we only draw one of them.}
\label{fig:weight_sectors}
\end{figure*}

Taken together, \eqref{eq:trace_rho} and \eqref{eq:weight-invariant} guarantee that at any point in the evolution, we can discard strings with weight $w(v) > N$ without incurring any approximation error: any string with \[w(v) > N\] will only produce other strings $v'$ with weight \[w(v') \geq w(v) > N.\]
We will refer to this strategy of discarding strings based on their weight as \emph{weight truncation}. The fact that weight truncation does not introduce any error in the expectation value is in stark contrast with Pauli and Majorana propagation, where weight truncation is a heuristic motivated by the small (or even vanishing) overlap of high-weight strings with typical initial states. Moreover, it introduces a ``backflow error'': high-weight strings can recombine into lower-weight ones \cite{DAnnaNysCarrasquilla2026, MillerEtAl2025MajoranaPropagation, rudolph2025paulipropagationcomputationalframework}. In the Dirac representation there is no such backflow, and the number of terms that can contribute to \eqref{eq:evolution-op-expansion} is upper-bounded by
\begin{equation}
  \sum_{q=0}^{N}\binom{M}{q}^2 \in \tilde{\mathcal{O}}\left(M^{2N}\right),
\end{equation}
and hence grows polynomially in $M$ for $N\in \mathcal{O}(1)$. For the case of dense filling, switching to the hole basis (where we replace $f^\dagger f$ with $f f^\dagger$) gives again a favorable scaling. For more details on this procedure, see Appendix \ref{app:hole_basis}.
Thus, the number of terms is bounded by
\begin{equation}
    T_{N} = \sum_{q=0}^{\min(N,M-N)}\binom{M}{q}^2 \in \tilde{\mathcal{O}}\left(M^{2\min(N,M-N)}\right).\label{eq:relevant-string-count}
\end{equation}

While it is true that weight truncation in MP reduces the number of strings to scale polynomially as well, the backflow from high weights is a distinctive feature of challenging MP simulations, and it was characterized for random unstructured circuits \cite{MillerEtAl2025MajoranaPropagation} only. On the other hand, DP shows the same reduction to polynomial scaling for $N\in \mathcal O(1)$, but the backflow is provably non-existent.

If we consider systems with multiple species of fermions, and the dynamics we want to simulate conserves each species separately, then the weight monotonicity condition holds for all species $j$ individually
\begin{equation} \label{eq:weight-invariant-species}
    w_j(U^\dagger GU) = w_j(G).
\end{equation}
Appendix \ref{sec:proof_weight_invariance} also contains a proof of this statement.

We now investigate the two efficient regimes of DP: Gaussian and Clifford circuits.

\subsection{Exact Gaussian and Clifford simulation}
\label{sec:gaussian-clifford}
\paragraph{Gaussian dynamics.}
We now show that in the Dirac formalism the expectation values of \emph{local observables} subject to number-conserving Gaussian unitary transformations are simulated efficiently without requiring approximations or truncation.

Consider a number-conserving Gaussian unitary generated by an
$M\times M$ Hermitian single-particle matrix $h=h^\dagger$. We define the
corresponding single-particle unitary matrix as
$V_h(\theta)=e^{\I\theta h}\in U(M)$, and the many-body generator and unitary
rotation as
\begin{equation}
  G(h)=\sum_{i,j=1}^{M}h_{ij}f_i^\dagger f_j,
  \qquad
  R_h(\theta)=e^{-\I\theta G(h)}.
  \label{eq:gaussian-definition}
\end{equation}
Because each term in $G(h)$ contains exactly one creation and one annihilation
operator, $R_h(\theta)$ preserves particle number.

Writing $V\equiv V_h(\theta)$, the Heisenberg action on a single creation
operator reads
\begin{equation}
  R_h^\dagger f_i^\dagger R_h
  =\sum_{j=1}^{M}V_{ji}f_j^\dagger.
  \label{eq:gaussian-linear}
\end{equation}
For an ordered index set $I=(i_1<\cdots<i_p)$, 
the insertion of $\mathbb I = R_h R_h^\dagger$ and expansion using \eqref{eq:gaussian-linear} gives
\begin{equation}
\begin{split}
    R_h^\dagger
  f_{i_1}^\dagger\cdots f_{i_p}^\dagger
  R_h & = \prod_{k=1}^p \left(\sum_{j_k=1}^M V_{j_k i_k} f^\dagger_{j_k}\right)\\
  & = \sum_{j_1,\cdots,j_p} V_{j_1i_1}\cdots V_{j_p i_p} (-1)^{s_{J}} f_{j_1}^\dagger \cdots f_{j_p}^\dagger\\
  & =
  \sum_{j_1<\cdots<j_p}
  \det(V_{J,I})
  f_{j_1}^\dagger\cdots f_{j_p}^\dagger,
  \end{split}
  \label{eq:gaussian-minors}
\end{equation}
where $s_J$ is the parity of the permutation that reorders the creation operators and $V_{J,I}$ is the submatrix with output rows $J$ and input columns $I$. Hence $p$ creation operators are mapped to a linear combination consisting only of operators with $p$ creation operators.
The corresponding annihilation-only formula is obtained by replacing $V$
with $V^*$. Therefore, Gaussian rotations preserve the numbers $p$ and $q$ of creation and 
annihilation operators in each string separately. Hence, every output
generated from an input string containing $p$ creation operators and $q$
annihilation operators has the same Dirac-weight $q$. The number of possible
canonical output strings is bounded by
\begin{equation}
  \binom{M}{p}\binom{M}{q}.
  \label{eq:gaussian-output-count}
\end{equation}
For the number-conserving strings considered here, it additionally holds that $p=q=w(v)$. This demonstrates that when the initial observable has a Dirac expansion with strings of sufficiently low weight, i.e., 
\begin{equation}
    O = \sum_{v:\, w(v)\leq k} c_v \eta(v),
\end{equation}
 the corresponding number of possible strings during Gaussian dynamics is $\sum_{i=0}^k \binom{M}{i}^2$ and can be simulated exactly without requiring truncation.

\paragraph{Clifford unitaries.} It was demonstrated in Ref.~\cite{DAnnaNysCarrasquilla2026} that Heisenberg picture simulations in the Majorana basis allow for the efficient simulation of two families of circuits: Gaussian and fermionic Clifford circuits. Having discussed Gaussian simulation above, we now want to investigate the Clifford unitaries. The splitting rule for Majorana propagation is conceptually more straightforward than the DP splitting rule~\eqref{eq:splitting-rule}: any string that anticommutes with the generator of a Majorana rotation is split into $2$ strings. Furthermore, for any Majorana rotation of any weight, we can always choose a rotation angle that makes the splitting a one-to-one mapping. In contrast, this is not the case in DP, as general Dirac-Clifford operations were shown to be operations that permute the indices and swap annihilation and creation operators \cite{MagoulasEvangelista2026}. Since we are restricting to number-preserving operators, only index permutations are allowed. Thus, the number-preserving Clifford operations are
\begin{equation}
  V_{ji}=e^{\I\phi_i}\delta_{j,\pi(i)},
\end{equation}
where $\pi(\cdot)$ is a permutation of the modes and $\phi_i$ is an arbitrary phase. These operations are a special case of the Gaussian transformation \eqref{eq:gaussian-linear}.
In particular, two types of Gaussian rotations generate all number-preserving Clifford operations. The first ones are hopping rotations at discrete angles that swap the modes at sites $p$ and $q$,
\begin{equation}
   S_{pq}= \exp\left[
     -\frac{\pi}{2}
     \left(f_p^\dagger f_q-f_q^\dagger f_p\right)
   \right].
 \end{equation}
The second family is given by the arbitrary single-mode phase rotations,
\begin{equation}
   D_p(\phi)=e^{-\I\phi n_p},
 \end{equation}
 where $\phi\in \mathbb{R}$.
Using that arbitrary two-site swaps (transpositions) generate the permutation group, we can conclude that $S_{pq}$ and $D_p(\phi)$ together generate the Clifford group for the number-preserving unitary operator space,
\begin{equation}
    \mathcal{C}_M = \langle \{S_{pq},D_p(\phi)|p,q\in[M], \phi\in \mathbb{R}\} \rangle.
\end{equation}
Therefore, the number-conserving Clifford update is an especially simple
 subcase of Gaussian propagation that remains efficient for Dirac strings of any weight.

\section{Experiments}\label{sec.experiments}
We demonstrate the performance of Dirac propagation on two simulations of dynamics following a quantum quench: (i) the $2$D Fermi-Hubbard model at different interaction strengths (Section \ref{sec:FHres}) and (ii) the lithium hydride diatomic molecule with progressively larger orbital sets (Section \ref{sec:LiHres}). As discussed below, we approximate the dynamics $\exp(-\I \delta \tau H)$ by means of a Trotter expansion and make use of additional truncation strategies beyond weight truncation discussed in Section \ref{sec:weight_monot}. 

\paragraph{Trotterization.} Both families of Hamiltonians we consider in this Section are of the form
\begin{equation}
H=\sum_v h_v (\eta(v) + \eta(v)^\dagger),
\end{equation}
and we aim at implementing the dynamics unitary $U(\delta \tau)= \exp(-\I \delta \tau H)$ for a short time interval $\delta \tau$ by a second-order Trotterization 
\begin{equation}\label{eq:trotter-second}
\begin{split}
    \exp(-\I \delta \tau H) \approx &\left(\prod_{l=1}^{L} \exp(-\I \theta_l (\eta(v_l)+\eta(v_l)^\dagger))\right)\cdot \\
    &\left(\prod_{l=L}^{1} \exp(-\I \theta_l (\eta(v_l)+\eta(v_l)^\dagger))\right),
    \end{split}
\end{equation}
where $\theta_l = \delta \tau h_l / 2$.

\paragraph{Weight truncation for different species.} Both families of Hamiltonians we consider host spinful fermions, and both of them are spin-preserving. For simulating a system of $N_{\uparrow}, N_{\downarrow}$ fermions under such spin-preserving Hamiltonians, \eqref{eq:weight-invariant-species} guarantees that we can perform weight truncation on the per-species weight $w_{\uparrow, \downarrow}$ separately, namely that we discard strings $\eta(v)$ with
\begin{equation}
    w_{\uparrow, \downarrow}(v) > N_{\uparrow, \downarrow}
\end{equation}
without incurring any approximation error. Notice that this is a more stringent truncation criterion than truncating strings with 
\[w(v) > N_{\uparrow} + N_{\downarrow}.\]
In this case, we find a tighter upper bound on the number of terms $T_{N_{\uparrow}, N_{\downarrow}}$ than the general case upper bound $T_N$ derived in \eqref{eq:relevant-string-count}
\begin{equation}\label{eq:relevant-string-count-spin}
    T_{N_{\uparrow}, N_{\downarrow}}= \prod_{\sigma=\uparrow,\downarrow}\sum_{q=0}^{\min(N_\sigma,M_\sigma-N_\sigma)}\binom{M_\sigma}{q}^2.
\end{equation}

\paragraph{Additional truncations.}
While weight truncation gives exact analytical guarantees, we introduce two additional truncation schemes to carry out the simulations in practice: coefficient truncation and unpaired truncation. The goal of such truncation schemes is to provide additional ``truncation knobs'' to further reduce the computational complexity of the simulation, at the cost of incurring an approximation error in the expectation value.

Coefficient truncation is a standard truncation scheme for both PP and MP \cite{rudolph2025paulipropagationcomputationalframework,DAnnaNysCarrasquilla2026}. We select a truncation threshold $\varepsilon$, and at any point of the simulation, we check the coefficients in the Dirac expansion \eqref{eq:evolved-observable}, and we discard strings with
\begin{equation}
    |c_v| < \varepsilon.
\end{equation}
In practice coefficient truncation greatly helps in reducing the number of strings in the Dirac expansion, but larger values of $\varepsilon$ lead to larger errors $\Delta$ in the expectation value. We derive an upper bound on such errors in Appendix \ref{sec:coeff_trunc_proof}, where we show that they are upper bounded by
\begin{align}
    \Delta \leq \sum_{l=0}^{L-1}\Delta_{l},
\end{align}
where $\Delta_l$ is a per-gate upper bound that is directly computable from the propagated expansion and the discarded terms.

Unpaired truncation relies on the notion of unpaired weight $w_\text{unpaired}$ of a string $\eta(v)$
\begin{equation}
    w_\text{unpaired}(v) = \sum_{k=1}^M \delta_{v_{2k-1} \neq v_{2k}}.
\end{equation}
The unpaired weight captures how many modes of the string have either $f^\dagger_k$ or $f_k$ as sole operators. It was introduced for MP in \cite{DAnnaNysCarrasquilla2026}, and it is a truncation specialized to expectation values involving a single Fock initial state $\ket{K}$ in the notation of \eqref{eq:fock-states}, as in this case 
\begin{equation}
    \langle K | \eta(v) | K \rangle =0 \quad \forall~\eta(v): w_\text{unpaired}(v)> 0.
\end{equation}
The intuition behind this truncation scheme is that while strings with many unpaired modes can in principle transform to be fully paired again, the coefficient resulting from this transformation becomes so small that it can be discarded.

\subsection{Lattice electrons}\label{sec:FHres}
We consider the $2$D Fermi-Hubbard model 
\begin{equation}\label{eq:FH}
H=-t \sum_{\langle i, j\rangle, \sigma \in\{\uparrow, \downarrow\}}\left(f_{i\sigma}^{\dagger} f_{j\sigma}+f_{j\sigma}^{\dagger} f_{i\sigma}\right)+U \sum_i n_{i \uparrow} n_{i \downarrow}.
\end{equation}
Throughout this section we set $\hbar=1$ and measure energies in units of the hopping $t$, so that times $\tau$ are in units of $1/t$. We simulate the evolution of an initial state $\ket{\psi_0}$ consisting of $D$ doublons along a stripe, following the setup of Ref.~\cite{Brezinova2024}. Let $K_D = \{i_1,\cdots, i_D\}$ be the index set with the positions of the doublons, then 
\begin{equation}
    \ket{\psi_0} = f^\dagger_{i_D\uparrow}f^\dagger_{i_D\downarrow}\cdots  f^\dagger_{i_1\uparrow}f^\dagger_{i_1\downarrow}\ket0.
\end{equation}
We consider a $6\times 6$ lattice with periodic boundary conditions, and study the effect of various interaction strengths. To this end, we simulate quenches subject to the Fermi-Hubbard Hamiltonian~\eqref{eq:FH} with $U/t\in \{0, 1, 8\}$, with the last one entering the strongly correlated regime where most classical methods struggle.

For free fermions, the dynamics exhibit exact revivals arising from the commensurate single-particle spectrum. This is due to the fact that the nearest-neighbor dispersion is 
\begin{equation}
\epsilon_{\mathbf{k}}=-2t(\cos k_x+\cos k_y), 
\end{equation}
with $k_{x,y}=2\pi m_{x,y}/L$.
For $L=6$ the allowed values are $\cos(k_{x,y}) \in \{1,\frac12,-\frac12,-1\}$, and all single-particle energies are integer multiples of the hopping strength $t$. This spectral commensurability implies that for the $U/t=0$ case we have an exact revival at $\tau_{\mathrm{rev}}=2\pi$, namely we have
\begin{equation}\label{eq:revival}
    e^{-\I H\tau_{\mathrm{rev}}}=1,
\end{equation}
and we expect our numerical simulations to exhibit this feature as well.

In the numerical simulations, we track the total number of doublons
\begin{equation}
    N_{d} = \sum_i n_{i \uparrow} n_{i \downarrow}
\end{equation}
over time.
We implement the second-order Trotterization \eqref{eq:trotter-second} with $\delta\tau=0.1$, and simulate the system until $\tau=12$. 
We consider two different quenches. In the first quench we consider a stripe with $D=2$, which is treatable with exact diagonalization (ED) at all interaction strengths, whereas we pick $D=6$ for  a second quench that is not tractable with ED in the interacting regime. In the interacting regime, we compare the DP results with MP results obtained with \texttt{MajoranaPropagation.jl} \cite{MPjl}. In all simulations we present, DP and MP runs are carried out until we reach the maximum memory of our available cluster (around 1 TB of memory), which accommodates Dirac and Majorana expansions with $\sim 10^{10}$ strings.

We simulate the quenches using DP at different coefficient truncations $\varepsilon = 10^{-3}, 3\times 10^{-4}, 10^{-4}$, and impose weight truncation for the two spin sectors separately, $N_{\uparrow,\downarrow}=D$, since the Fermi-Hubbard Hamiltonian \eqref{eq:FH} is spin-preserving. For the simulation with $D=6$ doublons, we additionally use the unpaired weight truncation.

\begin{figure*}
\centering
\includegraphics[width=\textwidth]{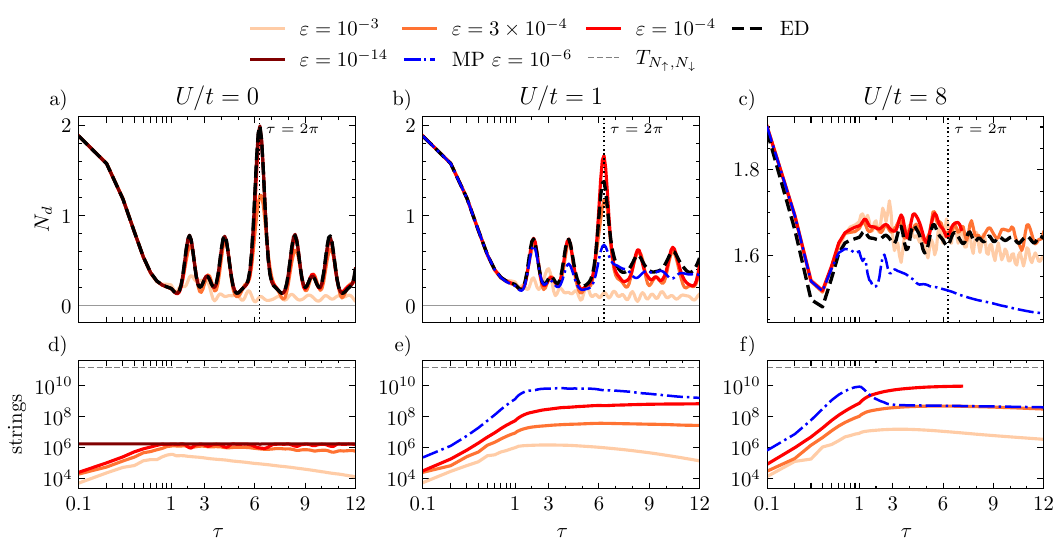}
\caption{
Number of doublons for a quench of $D=2$ doublons placed on adjacent sites of a $6\times6$ lattice, evolved under Fermi-Hubbard dynamics at different interaction strengths. (a) Comparison with ED in the free fermions case at $U/t = 0$, including a simulation with $\varepsilon=10^{-14}$.  (b)-(c) Comparison with both ED and MP in the interacting regime, for $U/t=1$ (b) and $U/t=8$ (c).
}
\label{fig:fh}
\end{figure*}

\paragraph{Results for $D=2$.}
The results for $U/t=0, 1, 8$ are displayed in Figure \ref{fig:fh}.
We simulate the Gaussian dynamics with $\varepsilon=10^{-3}, 3\times 10^{-4}, 10^{-4}, 10^{-14}$, the last one used to showcase that DP can simulate free fermions dynamics exactly. We observe the exact revival \eqref{eq:revival} of the state, illustrated by the peak in $N_d$ at $\tau = 2\pi$. 

At $U/t=1$ the DP simulations at  $\varepsilon=3\times 10^{-4}, 10^{-4}$ agree qualitatively with the ED results. In particular, the ED-predicted peaks in $N_d$ occur at the correct times. MP results ($\varepsilon=10^{-6}$) agree well with ED and DP at earlier times, but at later times they become very inaccurate and carry more strings than the most accurate DP simulation. MP significantly underestimates the revival peak.

At $U/t=8$ all three $\varepsilon=10^{-3}, 3\times 10^{-4}, 10^{-4}$ give results that agree qualitatively with ED, and we attribute the discrepancy at early times to Trotter error. We notice that in the strongly interacting regime, there is no signature of the exact revival that characterized the free fermions result. Due to energy conservation, dissociating a doublon requires transforming the large interaction energy into kinetic energy, and therefore the number of doublons tends to be stable for $\tau > 1$. On the other hand, MP deviates from ED and DP already at $\tau < 1$. Recent works \cite{facelli2026fastconvergencemajoranapropagation, zhao2026efficientclassicalsimulationweakly} show that the computational complexity of MP is connected to the non-Gaussianity of the simulation. This makes the number of strings in the simulation grow faster than in the $U/t=1$ case, and truncation strategies that both estimate the expectation value well and do not run out of memory on the cluster are hard to find in practice.

\begin{figure}
\centering
\includegraphics[width=.49\textwidth]{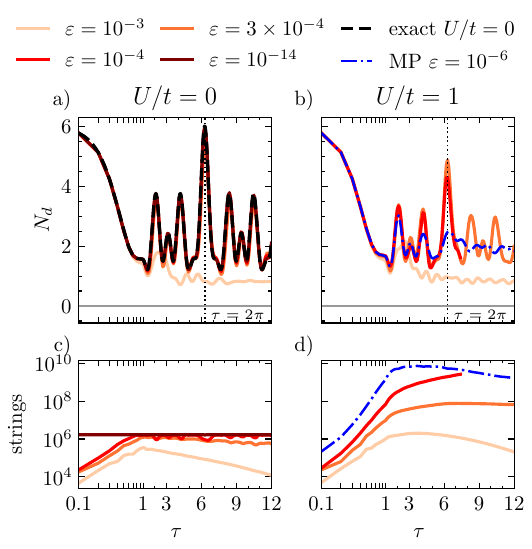}
\caption{Number of doublons for a quench of $D=6$ doublons placed on adjacent sites of a $6\times6$ lattice, evolved under Fermi-Hubbard dynamics at different interaction strengths. (a) Comparison with ED in the free fermions case at $U/t = 0$, including a simulation with $\varepsilon=10^{-14}$. (b) Comparison with MP in the interacting regime at $U/t=1$.}
\label{fig:fh_nd6}
\end{figure}

\paragraph{Results for $D=6$.}
The results for $U/t=0, 1$ are shown in Figure \ref{fig:fh_nd6}.
We simulate Gaussian dynamics with $\varepsilon=10^{-3}, 3\times 10^{-4}, 10^{-4}, 10^{-14}$, and as before we confirm that DP allows exact simulation of Gaussian dynamics. 
The results presented here for the interacting regime are additionally truncated with unpaired weight truncation of $6$. We observe that DP captures the revival peak at $\tau_\mathrm{rev}$ more accurately, whereas the MP simulation carries more strings than any DP simulation and completely misses the peak. For simulating a system with $6$ doublons, we notice that the upper bound \eqref{eq:relevant-string-count-spin} gives $T_{N_{\uparrow}, N_{\downarrow}}\approx 1.6\cdot 10^{25}$ strings, far beyond the possibilities of exact simulations.

\subsection{Isolated molecules}\label{sec:LiHres}

The polynomial scaling of DP is particularly favorable when the system has many fermionic modes but only a few fermions occupying them. Molecular systems are a natural candidate, since approaching the complete-basis-set limit means simulating a fixed number $N$ of electrons in an increasingly large orbital set with $M \gg N$ modes.

To illustrate this, we simulate lithium hydride, a standard benchmark for quantum simulations of quantum chemistry~\cite{kandala2017hardware,hempel2018quantum}, at fixed nuclear geometry. We approach the continuum limit by increasing the basis set while keeping the number of electrons fixed.

We consider a situation in which the Li nucleus sits at the origin and the H nucleus sits on the positive $z$ axis at $R = 1.60$~\AA . All 4 electrons are treated explicitly, without frozen core or active-space reduction. Throughout this section we use Hartree atomic units, $\hbar = m_e = e = 4\pi\varepsilon_0 = 1$, so that energies are in units of the Hartree $E_h$, lengths in units of the Bohr radius $a_0$, and times $\tau$ in units of $\hbar/E_h$. In a basis of $n$ real, orthonormal spatial orbitals $\{\phi_p\}$, the field-free electronic Hamiltonian reads~\cite{szabo1996modern,helgaker2000molecular}
\begin{equation}
  \begin{split}
  H_0 = {}& E_{\mathrm{nuc}} + \sum_{pq\sigma} h_{pq}\, f^\dagger_{p\sigma}f_{q\sigma} \\
  &+ \frac{1}{2}\sum_{pqrs}\sum_{\sigma\tau} (pq|rs)\,
f^\dagger_{p\sigma}f^\dagger_{r\tau}f_{s\tau}f_{q\sigma},
  \end{split}
  \label{eq:lih-h0}
\end{equation}
where $f^\dagger_{p\sigma}$ creates an electron of spin $\sigma\in\{\uparrow,\downarrow\}$
in orbital $\phi_p$, $h_{pq}$ are the one-electron integrals (kinetic energy and
nuclear attraction), $(pq|rs)$ are the electron-repulsion integrals, and $E_{\mathrm{nuc}}$ is the nuclear repulsion. 

The observable we simulate is the $z$ component of the total dipole moment, with origin at the Li nucleus,
\begin{equation}
  \hat\mu_z = Z_{\mathrm{H}} R - \sum_{pq\sigma} z_{pq}\, f^\dagger_{p\sigma}f_{q\sigma},
  \qquad z_{pq} = \langle\phi_p|z|\phi_q\rangle ,
  \label{eq:lih-dipole}
\end{equation}
with $Z_{\mathrm{H}} = 1$. All integrals are computed with
PySCF~\cite{sun2018pyscf,sun2020recent} in four Gaussian basis sets of
increasing size: STO-3G~\cite{hehre1969self,hehre1970self},
6-31G~\cite{ditchfield1971self,hehre1972self,dill1975self},
6-31G*~\cite{hariharan1973influence}, and
cc-pVDZ~\cite{dunning1989gaussian,prascher2011gaussian}. They give between $n = 6$ and $n = 19$ spatial orbitals, i.e.\ up to $38$ fermionic modes, see Table \ref{tab:lih-bases}.

\begin{table}[t]
  \caption{Basis sets for the LiH quench: number $n$ of spatial
    orbitals ($M=2n$ modes), number $N_g$
    of elementary gates per Trotter step, and squared overlap $|\langle\phi_0|\Phi_F\rangle|^2$ between the initial state $\ket{\Phi_F}$ and the exact ground state $\ket{\phi_0}$ of $H_0$.}
  \label{tab:lih-bases}
  \begin{ruledtabular}
  \begin{tabular}{lrrrc}
    Basis & $n$ & $M$ & $N_g$ & $|\langle \phi_0 |\Phi_F\rangle|^2 $ \\
    \hline
    STO-3G      &  6 & 12 & 956 & 0.9744\\
    6-31G       & 11 & 22 & 4\,940 & 0.9584\\
    6-31G*      & 16 & 32 & 13\,972 & 0.9545\\
    cc-pVDZ     & 19 & 38 & 17\,360 & 0.9452\\
  \end{tabular}
  \end{ruledtabular}
\end{table}

For the initial state $|\Phi_F\rangle$ we choose the restricted Hartree--Fock (RHF) Slater determinant of the molecule in a static, uniform electric field $F$ along $z$, i.e.\ the RHF solution of
\begin{equation}
  H_F = H_0 - F\hat\mu_z
  \label{eq:lih-hf}
\end{equation}
in the length gauge, obtained with the finite-field
method~\cite{cohen1965electric} with $F = 0.02$~a.u.

At $\tau = 0$ the field is switched off, and the state evolves under the
field-free Hamiltonian
\begin{equation}
\ket{\psi(\tau)} = e^{-\I H_0\tau}\ket{\Phi_F}. 
\end{equation}
We simulate the relaxation of the dipole,
\begin{equation}
  \langle\mu_z\rangle(\tau) = \langle \psi(\tau)|\hat\mu_z|\psi(\tau)\rangle
  \label{eq:lih-observable}
\end{equation}
up to $\tau_{\max} = 30$~a.u. Such field-free propagation of a prepared electronic state is a standard probe of real-time electronic-structure methods~\cite{krause2005timedependent,LiEtAl2020RealTime}. Since a single determinant is not an eigenstate of the interacting Hamiltonian, the dynamics contain both the relaxation of the field-induced polarization and correlation-driven electron motion, cf.\ Ref.~\cite{cederbaum1999ultrafast}.

\begin{figure*}
\centering
\includegraphics[width=\textwidth]{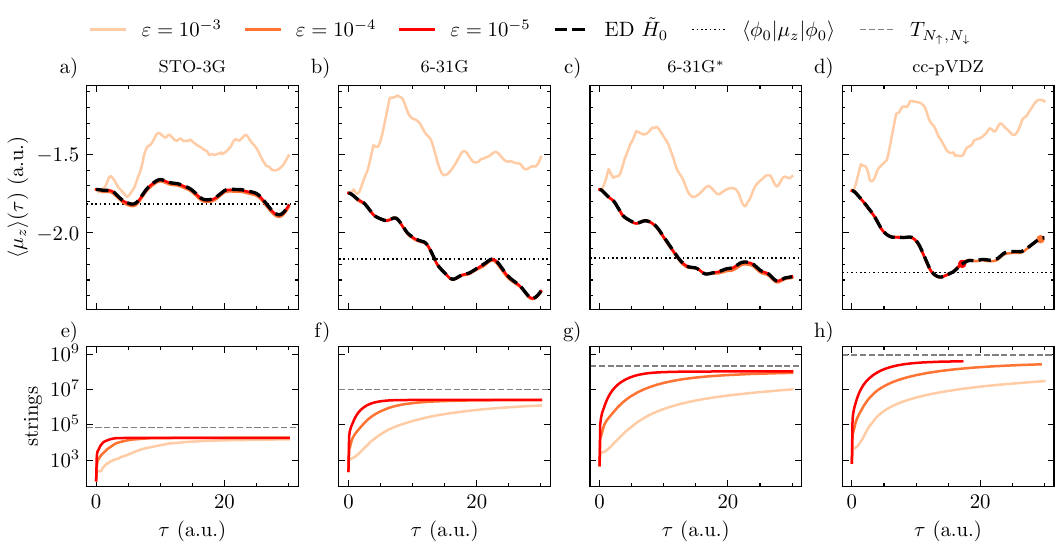}
\caption{Expectation value of the dipole moment $\mu_z$ of the HF ground state of a field-polarized Hamiltonian evolved under the zero-field Hamiltonian. (a)-(d) Dipole moment as a function of $\tau$ for different orbital basis sets and different DP coefficient truncations $ \varepsilon = 10^{-3}, 10^{-4}, 10^{-5}$. We also truncate the weights in both spin sectors, $N_{\uparrow, \downarrow}=2$. We compare the DP simulations with ED simulations of the truncated Hamiltonian $\tilde H_0$ used to build the second-order Trotterization, with $\delta\tau=0.1$. The dotted line marks the dipole moment $\langle\phi_0|\hat\mu_z|\phi_0\rangle$ in the exact ground state $\ket{\phi_0}$ of $H_0$. (e)-(h) Number of strings in each simulation. We observe that in all setups the number of strings becomes constant, and the total number of strings saturates below the upper bound \eqref{eq:relevant-string-count-spin}.}
\label{fig:lih}
\end{figure*}

We express $H_0$ and $\hat\mu_z$ in the canonical orbitals of the field-polarized RHF solution. In this basis the initial state is a single Fock state
\[\ket{\Phi_F}= f^\dagger_{1\uparrow}f^\dagger_{1\downarrow}f^\dagger_{2\uparrow}f^\dagger_{2\downarrow}|0\rangle,\]
and the field enters the dynamics only through the choice of orbitals: the propagation Hamiltonian $H_0$ contains no field term. Table~\ref{tab:lih-bases} also reports the squared overlap of $\ket{\Phi_F}$ with the exact ground state $\ket{\phi_0}$ of $H_0$, which exceeds $0.94$ for all orbital sets. The quench is therefore a weak perturbation of the ground state, and the dipole is expected to relax towards its ground-state value $\langle\phi_0|\hat\mu_z|\phi_0\rangle$, as we observe in Figure~\ref{fig:lih}.

To compile $e^{-\I H_0\delta \tau}$ into a Trotterization using fermionic gates, we impose a cutoff on the magnitude of the terms: we consider the truncated Hamiltonian $\tilde H_0$ where we keep terms from the field-polarized RHF solution with magnitude above a certain threshold $\lambda = 5\times 10^{-3}$. We then build the second-order Trotterization \eqref{eq:trotter-second} for $\tilde H_0$ with $\delta\tau=0.1$. In Table \ref{tab:lih-bases} we report the number of terms in a Trotter step for each orbital basis set we used. More efficient compilations of $e^{-\I H_0\delta\tau}$ exist, based on low-rank representations of the two-body term such as double factorization (DF)~\cite{Hohenstein2012Tensor,Parrish2012Tensor,Matsuzawa2020Jastrow,motta2021low} or tensor hypercontraction (THC)~\cite{Cohn2021Quantum,Oumarou2024acceleratingquantum}. Here our goal is only to illustrate that DP can serve as a simulation tool for molecular dynamics, and we leave DP simulations of DF- or THC-compiled circuits to future work.

We simulate the dynamics using DP with coefficient truncation $\varepsilon = 10^{-3}, 10^{-4}, 10^{-5}$ and weight truncation $N_{\uparrow,\downarrow}=2$. Figure \ref{fig:lih} shows the results for the different orbital sets. We compare the DP results with ED results of $\exp(-\I\tau H_0)$ and $\exp\nolimits(-\I\tau \tilde H_0)$. We observe that $\varepsilon = 10^{-4}, 10^{-5}$ follow the exact solution of $\tilde H_0$ at all times, whereas $\varepsilon = 10^{-3}$ gives completely wrong predictions. This is related to an interplay between $\lambda$ and coefficient truncation: in the splitting rule \eqref{eq:splitting-rule} new strings are generated at order $\sin(\theta) \approx \theta = h\delta \tau$, where $h$ is the coefficient of the corresponding term in the Hamiltonian. For many terms we hence have $\abs{h}\delta\tau < \varepsilon$, which means that terms are truncated as soon as they are created. The difference between the dynamics of $H_0$ and $\tilde H_0$ is also noticeable. Reducing $\lambda$ would recover the exact dynamics of the full Hamiltonian, but at the cost of much higher gate counts. 
We also observe that each simulation stabilizes at a certain number of strings, slightly below the upper bound \eqref{eq:relevant-string-count-spin}.

\section{Conclusions}\label{sec.conclusions}
We introduced Dirac propagation, a Heisenberg picture classical method for computing expectation values of $N$-particle states evolving under number-preserving unitaries. The method relies on the exact evolution of Dirac strings, followed by systematic truncation strategies to mitigate the growth of the number of strings. We show that truncating strings with weight above the target particle sector does not introduce errors in the expectation value. DP therefore shows no backflow from high to low weights, whereas other Heisenberg picture methods like PP or MP struggle to provide theoretical accuracy guarantees because of operator backflow. The absence of backflow implies that exact expectation values are computable with a number of strings that scales polynomially.
DP is also an exact simulator of number-preserving Gaussian dynamics, provided that the observables of interest are local (low weight) ones.

We apply DP to two different interacting fermionic systems: (i) a $2$D Fermi-Hubbard simulation, where we perform multiple quenches of a simple initial state, going from free fermions dynamics up to the strongly interacting regime. Thanks to the efficient scaling in the number of terms, we can accurately probe the dynamics for longer timescales than other Heisenberg picture methods, and we numerically observe the exact state revival after a full period (as predicted from free fermions analytical calculations), and discuss how this effect survives in the interacting regime. Additionally (ii), we use the favorable polynomial scaling at fixed $N$ of DP to simulate quenches of a molecular wave function. Here, we consider the dipole moment of a LiH molecule, and we compute its expectation values for quenches performed in multiple basis sets with an increasing number of orbitals.
Together, these results show that DP enables controlled, systematic classical estimation of expectation values of local fermionic observables under number-preserving evolutions.

The structure of DP suggests several directions for future research. In quantum chemistry, its treatment of Gaussian orbital rotations and density-density interactions could be combined with low-rank Hamiltonian factorizations and particle-hole representations to further reduce computational costs. 
In variational settings, using DP to simulate candidate number-preserving circuits could improve classical state preparation algorithms. 
Other opportunities include applications to correlated and thermal initial states, the calculation of dynamical correlation functions, and hybrid Schrödinger–Heisenberg simulation strategies. 
The weight monotonicity of DP allows lower weight sectors to be simulated independently of higher ones. In practice, the memory bottleneck of propagation methods could be alleviated by simulating one weight sector after the other starting from the lowest one.
Beyond fermionic systems, the underlying ideas could motivate new operator propagation schemes for other number-conserving models, including spin systems formulated in terms of $S_\pm$ operators and few-excitation bosonic systems. Finally, DP offers a promising framework for the classical simulation and benchmarking of emerging fermionic quantum simulators.

\section*{ACKNOWLEDGMENTS}
APS acknowledges
the Swiss National Science Foundation for its support
through the grant TMPFP2\_234085. 

\section*{AI usage statement}
The authors acknowledge the use of generative AI tools (Claude, ChatGPT) for the implementation of the code, verification of the main results and literature review. The authors reviewed, verified, and revised all AI-assisted content and take full responsibility for the content of this work.

\clearpage  

\appendix 

\section{Weight invariance under number-conserving evolution}\label{sec:proof_weight_invariance}
The Dirac-weight defined in Eq.~\eqref{eq:weight} can be understood as the smallest particle number on which a Dirac expansion acts nontrivially. To demonstrate this, we first show the following.
\begin{lemma}\label{lem:weight-sector-vanish}
A number-conserving Dirac expansion $G\ne0$ with weight $w(G) = k$ annihilates every state in $\mathcal F_r$ for $r<k$, i.e.,
\begin{equation}
  G \ket{\psi} = 0, \quad \forall \ket{\psi} \in \mathcal F_r,
\end{equation}
where $r < k$.
\end{lemma}

\begin{proof}
This follows from Definition \ref{def:weight-string} of the weight $w$ for a canonical string as the number of annihilation operators. Consider a canonical string $\eta(v)$ with $w(v)=k$. Moving all annihilation operators to the right introduces only a sign, since every exchanged pair belongs to distinct modes. The resulting product of $k$ annihilation operators
annihilates every state with fewer than $k$ particles.
Consider now a Dirac expansion $G = \sum_{v | c_v \neq 0} c_v \eta(v)$ with a weight $k=w(G)$. From Definition \ref{def:weight-expansion} for the weight of Dirac expansions, we find that $G$ annihilates all states in $\mathcal F_r$ for $r < k$.
\end{proof}

This leads to the following Lemma.
\begin{lemma}\label{lem:weight-sector-exists}
For a number-conserving Dirac expansion $G\ne0$,
\begin{equation}
  w(G)=\min\bigl\{r \mid \exists\,\ket{\psi}\in \mathcal F_r \text{ such that }G\ket{\psi} \ne 0 \bigr\}.
\end{equation}
\end{lemma}

\begin{proof}
Let $k=w(G)$. By Lemma~\ref{lem:weight-sector-vanish}, $G$ vanishes on every $\mathcal F_r$ with $r<k$. It remains to show that $G$ acts nontrivially on $\mathcal F_k$.

Strings of weight greater than $k$ (that might be present in $G$) vanish on $\mathcal F_k$ by Lemma~\ref{lem:weight-sector-vanish}. For a weight-$k$ string $\eta(v)$ in $G$, let $I$ be its set of annihilation
indices and $J$ its set of creation indices. Write $\ket{I}$ and $\ket{J}$ for the occupation states with precisely those modes occupied. Since $|I|=k$, the only $k$-particle occupation state on which $\eta(v)$ acts nontrivially is $\ket{I}$, and
\begin{equation}
    \eta(v)\ket{I}=\pm\ket{J}.
\end{equation}
Thus, restricted to $\mathcal F_k$, this string acts as $\pm\ket{J}\bra{I}$. Distinct canonical strings have distinct pairs $(I,J)$ and therefore contribute to distinct matrix
elements. Since at least one weight-$k$ coefficient is nonzero, these contributions cannot all cancel. Hence $G$ acts nontrivially on $\mathcal F_k$.
\end{proof}

\begin{proposition}[Weight invariance]
\label{prop:weight-invariance}
Consider a number-conserving Dirac expansion $G$. For any number-conserving unitary $U$
\begin{equation}
    w(U^\dagger G U)=w(G).
\end{equation}
\end{proposition}

\begin{proof}
The statement is trivial for $G=0$. Let $k=w(G)$ for $G \neq 0$. By Lemma~\ref{lem:weight-sector-vanish}, $G$ annihilates every state in $\mathcal F_r$ for $r<k$. Since $U$ preserves each particle-number subspace, $U\ket{\psi}\in \mathcal F_r$ whenever $\ket{\psi}\in \mathcal F_r$. Hence
\begin{equation}
    U^\dagger G U\ket{\psi}=0 \qquad\text{for every }\ket{\psi}\in \mathcal F_r,\quad r<k.
\end{equation}

By Lemma~\ref{lem:weight-sector-exists}, there is a state $\ket{\psi}\in \mathcal F_k$ with $G\ket{\psi}\neq0$. Define $\ket{\phi}=U^\dagger\ket{\psi} \in \mathcal F_k$. Then
\begin{equation}
    U^\dagger G U\ket{\phi}
    =U^\dagger G\ket{\psi}\neq0,
\end{equation}
because a unitary operator maps a nonzero vector to a nonzero vector. Hence, we have found a state $\ket{\phi} \in \mathcal F_k$ on which $U^\dagger G U$ acts nontrivially. This shows that the smallest particle number on which $U^\dagger G U$ acts nontrivially is exactly $k$, proving the claim.
\end{proof}
We note that this proof extends the findings of \cite{MagoulasEvangelista2026}, where it was shown that Dirac-Clifford transformations cannot alter the weight of Dirac strings.

\subsection{Different fermionic species}
In the case of simulating the dynamics of $s$ fermionic species that are all individually preserved by the dynamics, Lemmata \ref{lem:weight-sector-vanish}, \ref{lem:weight-sector-exists} can be applied to each species separately.
\begin{lemma}\label{lem:species-weight-sector-vanish}
A number-conserving Dirac expansion $G\ne0$ with $j$-weight $w_j(G) = k$ annihilates every state in $\mathcal F_{N_1,\cdots, N_j,\cdots,N_s}$ for $N_j<k$, i.e.\ 
\begin{equation}
  G \ket{\psi} = 0, \quad \forall \ket{\psi} \in \mathcal F_{N_1,\cdots, N_j,\cdots,N_s},
\end{equation}
where $N_j < k$.
\end{lemma}
\begin{proof}
The proof of Lemma \ref{lem:species-weight-sector-vanish} follows directly from the proof of Lemma \ref{lem:weight-sector-vanish}, by the replacement $w \to w_j$.
\end{proof}
\begin{lemma}\label{lem:species-weight-sector-exists}
For a number-conserving Dirac expansion $G\ne0$,
\begin{equation}
  w_j(G)=\min\bigl\{r \mid \exists\,\ket{\psi}\in \mathcal M_r^{(j)}: G\ket{\psi} \ne 0 \bigr\},
\end{equation}
for $\mathcal M_r^{(j)}$ the Fock space restricted to states with $r$ particles of species $j$, and any number of particles in all other species,
  \begin{equation}\label{eq:Mj}
      \mathcal M_r^{(j)} = \bigoplus_{N_1=0}^{M_1}\cdots \bigoplus_{N_{j-1}=0}^{M_{j-1}}\bigoplus_{N_{j+1}=0}^{M_{j+1}}\cdots \bigoplus_{ N_s=0}^{M_s}  \mathcal F_{N_1,\cdots, r,\cdots, N_s}
  \end{equation}
\end{lemma}
\begin{proof}
Let $k_j=w_j(G)$. By Lemma \ref{lem:species-weight-sector-vanish} we have 
\begin{equation}
    G \ket\psi = 0 \quad \forall \ket\psi \in \mathcal M_{r < k_j}^{(j)}.
\end{equation}
Strings in $G$ with $j$-weight higher than $k_j$ will vanish on $\mathcal M_{k_j}^{(j)}$. We hence have to show that strings of $j$-weight $k_j$ act nontrivially on $\mathcal M_{k_j}^{(j)}$.

For a string $\eta(v)$ with $j$-weight $k_j$, let $I_j$ be its set of annihilation
indices and $J_j$ its set of creation indices of modes for the species $j$. Write $\ket{I_j; K}$ and $\ket{J_j; K}$ for the occupation states with precisely those species-$j$ modes occupied, and the occupation of the modes of all other species specified by $K$. We have $\ket{I_j;K},\ket{J_j;K} \in \mathcal M_{k_j}^{(j)}$ by construction and the definition \eqref{eq:Mj}. Independent of the ordering of the fermionic species and modes, we can write
\begin{equation}
    \eta(v) = \sigma \prod_{l=1}^s \eta(v_l)
\end{equation}
for $\sigma=\pm1$ a reordering sign and $v_l$ the Dirac string containing only the modes of species $l$ that are present in $v$. We can also exchange the order of the $\eta(v_j)$ at the price of possible reordering signs, hence we have
\begin{equation}
    \eta(v) \ket{I_j; K} = \pm\sigma \prod_{\substack{l=1\\l\neq j}}^s \eta(v_l) \ket{J_j; K}.
\end{equation}
This procedure can be repeated for any species, and hence there exists a $K$ so that 
\begin{equation}
    \prod_{\substack{l=1\\l\neq j}}^s \eta(v_l) \ket{J_j; K} \neq 0.
\end{equation}
\end{proof} 
\noindent Since Lemmata \ref{lem:species-weight-sector-vanish} and \ref{lem:species-weight-sector-exists} hold, we can formulate the per-species version of Proposition \ref{prop:weight-invariance}:
\begin{proposition}[Weight invariance for individual fermionic species]
\label{prop:species-weight-invariance}
Consider a number-conserving Dirac expansion $G$ and  a unitary $U$ that is number-conserving for all the fermionic species $j=1,\cdots,s$. Then for $j=1,\cdots,s$ it holds that
\begin{equation}\label{eq:species-weight-invariant}
    w_j(U^\dagger G U)=w_j(G).
\end{equation}
\end{proposition}
\begin{proof}
Let $j=1,\cdots,s$. Let $k_j=w_j(G)$. By Lemma \ref{lem:species-weight-sector-vanish} any state $\ket\psi\in\mathcal M_{r<k_j}^{(j)}$ is annihilated by $G$. Since $U$ preserves all the different species, for $\ket\psi \in \mathcal F_{N_1,\cdots, r,\cdots,N_s}$ we have $U\ket\psi \in \mathcal F_{N_1,\cdots, r,\cdots,N_s}$ as well, which also implies $\ket\psi \in \mathcal M_{r}^{(j)}, U\ket\psi \in \mathcal M_{r}^{(j)}$. Hence for such states with $r<k_j$ 
\begin{equation}
    U^\dagger G U \ket\psi = 0.
\end{equation}
Let $\ket\psi \in \mathcal M_{k_j}^{(j)}$ be a state such that $G\ket\psi\neq 0$ (Lemma \ref{lem:species-weight-sector-exists}). Define
\begin{equation}
    \ket\phi = U^\dagger \ket\psi \in \mathcal M_{k_j}^{(j)}.
\end{equation}
Then 
\begin{equation}
    U^\dagger G U \ket\phi = U^\dagger G \ket\psi \neq 0,
\end{equation}
which proves that $U^\dagger G U$ acts nontrivially on $\mathcal M_{k_j}^{(j)}$, and hence \eqref{eq:species-weight-invariant} follows.

\end{proof}

\subsection{Weight invariance and time-reversal}

Weight monotonicity raises the question of time-reversal symmetry: for example, how is it possible that applying a unitary $U$ and then $U^\dagger$ gives back the original operator if the weight of the strings will always be non-decreasing?

In Pauli and Majorana propagation, weight  decreases when the two strings being multiplied share some sites in common, as Pauli and Majorana operators square to the identity, e.g. $\sigma_x^2=1$. For Dirac strings, on the other hand, we have $f^2=0$. As exemplified in \eqref{eq:splitting_example_commutators}, this property causes some strings in the splitting rule to be removed as they become $0$.

But this alone is insufficient, and an example best illustrates the other mechanism that restores time reversal.
Consider the case of propagating the hopping operator ($i<k$ in the canonical ordering) $f^\dagger_i f_k - f_i f^\dagger_k$ through a density-density interaction $U_{nn} = \exp(-\I\theta n_in_j)$. Using \eqref{eq:splitting_example} we find 
\begin{equation}
\begin{split}
    U_{nn}^\dagger (f^\dagger_i f_k - f_i f^\dagger_k) U_{nn} & = f_i^\dagger f_k + (e^{\I\theta}-1) f_i^\dagger n_j f_k \\ & - f_i f^\dagger_k - (e^{-\I\theta}-1) f_i n_j f_k^\dagger
\end{split}
\end{equation}
and applying the inverse evolution $\exp(+\I\theta n_in_j)$ gives us
\begin{widetext}
\begin{equation}
    \begin{split}
        &f^\dagger_i f_k - f_i f^\dagger_k = U_{nn}U_{nn}^\dagger (f^\dagger_i f_k - f_i f^\dagger_k) U_{nn}U_{nn}^\dagger\\
        & = U_{nn}(f_i^\dagger f_k + (e^{\I\theta}-1) f_i^\dagger n_j f_k 
        \quad- f_i f^\dagger_k - (e^{-\I\theta}-1) f_i n_j f_k^\dagger)U_{nn}^\dagger \\
        & = (f_i^\dagger + (e^{-\I\theta}-1)f_i^\dagger n_j) f_k + (e^{\I\theta}-1)(f_i^\dagger + (e^{-\I\theta}-1) f_i^\dagger n_j)n_j f_k-(f_i + (e^{\I\theta}-1)f_i n_j)f_k^\dagger- (e^{-\I\theta}-1)(f_i+(e^{\I\theta}-1)f_i n_j) n_j f_k^\dagger \\
        & = f_i^\dagger f_k + (e^{-\I\theta}-1 + e^{\I\theta} -1 + (e^{-\I\theta}-1)(e^{\I\theta} -1)) f_i^\dagger n_j f_k - f_i f_k^\dagger - (e^{\I\theta}-1 +e^{-\I\theta}-1 + (e^{\I\theta}-1)(e^{-\I\theta}-1)) f_i n_j f_k^\dagger \\
        &= f^\dagger_i f_k - f_i f^\dagger_k.
    \end{split}
\end{equation}
\end{widetext}
We hence see that all newly created terms are killed by exact cancellations (in this case, by $e^{\I\theta}-1 +e^{-\I\theta}-1 + (e^{\I\theta}-1)(e^{-\I\theta}-1) = 0$), and the only surviving term is the original hopping term.

\section{Extended methodology}

\subsection{Multiplication}
\label{app:multiplication}
Consider two Dirac strings $\eta(u)$ and $\eta(v)$,
\begin{align}
  \eta(u)&=\prod_{k=1}^{M}(f_k^\dagger)^{u_{2k-1}}(f_k)^{u_{2k}},\\
  \eta(v)&=\prod_{k=1}^{M}(f_k^\dagger)^{v_{2k-1}}(f_k)^{v_{2k}}.
\end{align}
The multiplication between the two strings is computed by the following steps:

\begin{enumerate}
    \item Reordering: First, we have to reorder the modes from the two strings such that they are next to each other.
    \begin{equation}
        \eta(u)\eta(v)=(-1)^{v^T \omega_U u}\prod_{k=1}^{M}(f_k^\dagger)^{u_{2k-1}}(f_k)^{u_{2k}}(f_k^\dagger)^{v_{2k-1}}(f_k)^{v_{2k}}
    \end{equation}
    where $\omega_U=\delta_{\lfloor j/2\rfloor>\lfloor i/2\rfloor}$ is a $2M \times 2M$ upper triangular matrix ($\lfloor \cdot\rfloor$ is the floor function)
\begin{equation}
    \omega_U=\begin{pmatrix}
        0&0&1&1&1&1 &\cdots& 1&1\\
        0&0&1&1&1&1 &\cdots& 1&1\\
        0&0&0&0&1&1 &\cdots& 1&1\\
        0&0&0&0&1&1 &\cdots& 1&1\\
        \vdots & \vdots &\vdots &\vdots &\vdots &\vdots &\ddots  & \vdots & \vdots \\
        
        0&0&0&0&0&0 &\cdots& 1&1\\
        0&0&0&0&0&0 &\cdots& 1&1\\
        
        0&0&0&0&0&0 &\cdots& 0&0\\
        0&0&0&0&0&0 &\cdots& 0&0\\
    \end{pmatrix}
\end{equation}
    \item Contraction and hole mode flag: Recall the multiplication table from \eqref{eq:multiplication_table} that we now translate into index-wise rules for each mode
    \begin{equation}
        \renewcommand{\arraystretch}{1.15}
        \setlength{\arraycolsep}{4pt}
        \begin{array}{cc|cccc}
         & & \multicolumn{4}{c}{
           (v_{2k-1},v_{2k})
         }\\
         & & 00\;(\id) & 01\;(f) & 10\;(f^\dagger) & 11\;(n)\\
        \cline{2-6}
        \multirow{4}{*}{
          \rotatebox[origin=c]{90}{
            $(u_{2k-1},u_{2k})$
          }
        }
         & 00\;(\id)       & 00 & 01 & 10    & 11\\
         & 01\;(f)         & 01 &  \cross & 00-11 & 01\\
         & 10\;(f^\dagger) & 10 & 11 &   \cross   & \cross \\
         & 11\;(n)         & 11 & \cross  & 10    & 11
        \end{array}
        \label{eq:local-product}
    \end{equation}
    Let us define an auxiliary bit string $z \in \{0,1\}^{2M}$ that allows us to describe the mode contraction process. Let us also define an index set $I$ that stores the common modes, i.e.,
\begin{equation}
    I = \{k: (u_{2k-1}|u_{2k})=(v_{2k-1}|v_{2k})\}
\end{equation}
where $|$ is the bit ``OR'' operation. For $k\notin I$, the bit contraction is stored in $z$ as
\begin{equation}
    z_{2k-1}=u_{2k-1}\oplus v_{2k-1} \quad z_{2k}=u_{2k}\oplus v_{2k}.
\end{equation}
We now sequentially move through every common mode $k \in I$ and compute the contraction. We distinguish between the following three different situations
\begin{itemize}
    \item Deletion: If at any point, $u_{2k-1}u_{2k}^c v_{2k-1}+u_{2k}v_{2k-1}^c v_{2k}=1$, where $x^c\equiv 1-x$ denotes the bit complement,
    \begin{equation}
        \eta(u)\eta(v)=0
    \end{equation}
    \item Hole flag: We also have to flag the modes that result in the hole mode $f f^\dagger$. We store these modes in an index set $F$. That means
    \begin{equation}
        F=\{k \in I: u_{2k-1}^c u_{2k} v_{2k-1} v_{2k}^c=1\}
    \end{equation}
    \item Other cases: For every other case, the resultant bit contraction can be safely stored in the auxiliary string $z$,
    \begin{equation}
        z_{2k-1}= u_{2k-1}+u_{2k}^c v_{2k-1} \quad z_{2k}= u_{2k} v_{2k-1}^c+ v_{2k}
    \end{equation}
\end{itemize}

\item Conversion to canonical string basis: In the final step, we use the hole mode flag $F$ to convert $f f^\dagger$ to $\id -f^\dagger f$. 
\begin{equation}
    \eta(u)\eta(v) = (-1)^{v^T \omega_U u} \prod_{k \in F^c}(f_k^\dagger)^{z_{2k-1}}(f_k)^{z_{2k}}\prod_{k \in F}(\id - f^\dagger_k f_k)
\end{equation}
where $F^c=[M]\backslash F $. Thus, the multiplication of two Dirac strings will result in a Dirac expansion with $2^{|F|}$ strings.
\end{enumerate}

\subsection{Hermitian conjugation}
Let us define an index set $L$ that stores the unpaired modes in a Dirac string $\eta(v)$, i.e.,
\begin{equation}
    L = \{k: v_{2k-1} \oplus v_{2k}=1 \}.
\end{equation}
For the Hermitian conjugate of a canonical Dirac string, we find
\begin{equation}
  \eta(v)^\dagger=(-1)^{|L|(|L|-1)/2}\eta(v^{\mathrm{hc}}),
  \label{eq:string-adjoint}
\end{equation}
where $v^{\mathrm{hc}}$ is the binary vector obtained from $v$ by swapping the two bits at every mode,
\begin{equation}
  (v^{\mathrm{hc}})_{2k-1}=v_{2k},
  \qquad
  (v^{\mathrm{hc}})_{2k}=v_{2k-1}.
  \label{eq:hc-word}
\end{equation}
This bit swap accounts for $f_k\leftrightarrow f_k^\dagger$. For example, if $v=1001$, then $\eta(v)=f_1^\dagger f_2$ and
\begin{equation}
  (f_1^\dagger f_2)^\dagger=\eta(1001)^\dagger
  =-\eta(0110)
  =-f_1f_2^\dagger.
  \label{eq:string-adjoint-example}
\end{equation}
The two bit pairs are swapped locally, and reordering the unpaired modes gives the minus sign. Hermitian conjugation changes the weight to $w(v^{\mathrm{hc}})=\sum_k v_{2k-1}$. It therefore preserves the weight of a number-conserving string, but not necessarily that of a general string.

\subsection{Normal ordering}\label{app:normal-ordering}
While the binary representation above is useful for computations and implementation of DP, we also want to rewrite strings in the standard \emph{normal-ordered} form, which is more suitable for theoretical analysis. Let $I,J,K \subseteq [M]$ be disjoint index sets. We define
\begin{align}    
\mathbf{f}_I^\dagger&=f_{i_1}^\dagger\cdots f_{i_q}^\dagger,\\
\mathbf{f}_J&= f_{j_q}\cdots f_{j_1}\\
\mathbf{n}_K&= n_{k_1} \ldots n_{k_r}\\
\mathbf{n}_K\mathbf{f}_I^\dagger \mathbf{f}_J
&= n_{k_1} \ldots n_{k_r} f_{i_1}^\dagger \cdots f_{i_q}^\dagger
f_{j_q}\cdots f_{j_1},
\end{align}
where ($i_1<\ldots <i_q \in I$), ($j_1<\ldots <j_q \in J$) and ($k_1<\ldots <k_r \in K$). Density operators and creation operators are thus ordered by increasing mode index and annihilation operators by decreasing mode index. Notice that we take $|I| = |J|$ since we consider particle-preserving operators. The canonical ordering and normal ordering for the same occupied modes differ only by a sign. One can show that
\begin{equation}
    \mathbf{n}_K\mathbf{f}_A^\dagger \mathbf{f}_B= (-1)^{v_B^T \omega_U v_A+\frac{|B|(|B|-1)}{2}} \eta(v)
\end{equation}
where
\begin{align}
    A &= \{k: v_{2k-1}=1, v_{2k}=0\}\\
    B &= \{k: v_{2k-1}=0, v_{2k}=1\}\\
    K &= \{k: v_{2k-1}=1, v_{2k}=1\}\\
    v_A &= \text{Dirac string associated with } \mathbf{f}^\dagger_A \\
    v_B &= \text{Dirac string associated with } \mathbf{f}_B
\end{align}
 We choose this representation because it makes it easier to derive the Hilbert--Schmidt inner product for operators. One can show that \cite{Bach_2019}
\begin{align}
    \langle \textbf{n}_K \textbf{f}^\dagger_A \textbf{f}_B \,|\,\textbf{n}_L \textbf{f}^\dagger_C \textbf{f}_D\rangle_{HS} &= \Tr(\textbf{f}_B^\dagger \textbf{f}_A \textbf{n}_K \textbf{n}_L \textbf{f}^\dagger_C \textbf{f}_D)\\
    &=\delta_{A,C}\delta_{B,D} 2^{M-|A \cup B \cup L \cup K|}.\label{orthogonality}
\end{align}
The expressions can be found by the following counting argument. We get $A=C$ and $B=D$ for a non-zero inner product as any other case would give a zero overlap with every Fock state. Further, for a non-zero overlap, the sites occupied by $A$ must be empty while the sites occupied by $(B \cup L\cup K)$ must be filled with fermions. 
One can also show that
\begin{equation}
    \Tr_N(\textbf{f}_B^\dagger \textbf{f}_A \textbf{n}_K \textbf{n}_L \textbf{f}^\dagger_C \textbf{f}_D)= \delta_{A,C}\delta_{B,D} \binom{M-|A\cup B \cup L \cup K|}{N-|B \cup L \cup K|}
\end{equation}
Equation \eqref{orthogonality} points to a direct sum structure based on unpaired operators (i.e., $A$ and $B$ in $\textbf{n}_K \textbf{f}^\dagger_A \textbf{f}_B$).

We give a few examples to illustrate this. Consider a single-site system. The space of balanced strings is spanned by $\{\id,f^\dagger_1 f_1\}$. If we form an inner product matrix (Gram matrix) for such a basis, we get the following

\begin{equation}
G=\begin{array}{c}
\\
\id \\
f_1^{\dagger} f_1
\end{array}
\begin{array}{cc}
\id & f_1^{\dagger} f_1 \\
\left(\begin{array}{c} 2\\1 \end{array}\right. &
\left.\begin{array}{c} 1\\1 \end{array}\right)
\end{array}
\end{equation}

Consider a two-site system. The space of balanced strings is spanned by $\{\id, f_1^\dagger f_1, f_2^\dagger f_2, f_1^\dagger f_1 f_2^\dagger f_2, f_1^\dagger f_2, f_2^\dagger f_1 \}$.

\begin{equation}
G=\begin{array}{c}
\\
\id \\
f_1^{\dagger} f_1 \\
f_2^{\dagger} f_2 \\
f_1^{\dagger} f_1 f_2^{\dagger} f_2 \\
f_1^{\dagger} f_2 \\
f_2^{\dagger} f_1
\end{array}
\begin{array}{cccccc}
\id & f_1^{\dagger} f_1 & f_2^{\dagger} f_2 & f_1^{\dagger} f_1 f_2^{\dagger} f_2 & f_1^{\dagger} f_2 & f_2^{\dagger} f_1 \\
\left(\begin{array}{c} 4\\2\\2\\1\\0\\0 \end{array}\right. &
\begin{array}{c} 2\\2\\1\\1\\0\\0 \end{array} &
\begin{array}{c} 2\\1\\2\\1\\0\\0 \end{array} &
\begin{array}{c} 1\\1\\1\\1\\0\\0 \end{array} &
\begin{array}{c} 0\\0\\0\\0\\1\\0 \end{array} &
\left.\begin{array}{c} 0\\0\\0\\0\\0\\1 \end{array}\right)
\end{array}
\end{equation}
Effectively, $G = G_0 \oplus G_{f_1^\dagger f_2} \oplus G_{f_2^\dagger f_1}$. Notice further that the matrices $G_0, G_{f_1^\dagger f_2}, G_{f_2^\dagger f_1}$ share the same structure,

\begin{equation}
    G_0 = \begin{pmatrix}
        2 & 1\\
        1 & 1
    \end{pmatrix}^{\otimes 2}
\end{equation}
\begin{equation}
    G_{f_1^\dagger f_2} = G_{f_2^\dagger f_1} = \begin{pmatrix}
        2 & 1\\
        1 & 1
    \end{pmatrix}^{\otimes 0}
\end{equation}

We can now extend this to any number of sites. Given $M$ sites, the Gram matrix $G$ of the balanced strings is given as 

\begin{equation}
    G = \bigoplus_{\substack {A, B \subseteq[M]\\ A\cap B = \emptyset}} G_{\textbf{f}^\dagger_A \textbf{f}_B}
\end{equation}
where 
\begin{equation}
    G_{\textbf{f}^\dagger_A \textbf{f}_B} = \begin{pmatrix}
        2 & 1\\
        1 & 1
    \end{pmatrix}^{\otimes (M-2|A|)}
\end{equation}
Within each block, we remove the $(A \cup B)$ sites and order the row and column indices in increasing site index and increasing weight.
Now that we have an efficient way of constructing the Gram matrix in normal ordering, we can find the Gram matrix coefficients in the canonical ordering by a simple label matching (the sign will not play a role).

We can also find the inner product between two Dirac expansions. Let $R = \sum_u c_u \eta(u)$ and $S = \sum_{v} d_v \eta(v)$ be two Dirac expansions. The inner product between $R$ and $S$ is given as 

\begin{align}
    \langle R \, | \, S \rangle_{HS} &= \sum_{u, v} c_u^* d_v \Tr(\eta(u)^\dagger \eta(v))\\
    &= \sum_{u, v} c_u^* d_v G_{u v}.
\end{align}

\subsection{Hole basis}\label{app:hole_basis}
Recall that we refer to $\mathcal D_1$ as the local canonical Dirac operator basis
\begin{equation}
    \mathcal D_1=\{\id,f,f^\dagger,n=f^\dagger f\}.
\end{equation}
An equally valid choice of local basis could have been
\begin{equation}
    \mathcal D_1^h=\{\id,f,f^\dagger,h=ff^\dagger\}
\end{equation}
which we refer to as the local hole basis. One can now define a hole Dirac string as
\begin{equation}
  \eta^h(v)=(f_1)^{v_1}(f^\dagger_1)^{v_2} \cdots (f_M)^{v_{2M-1}}(f^\dagger_{M})^{v_{2M}},
  \label{eq:eta_h}
\end{equation}
indexed by the $2M$-bitstring $v \in \{0, 1\}^{2M}$. This allows us to describe the \textit{hole operator basis} 
\begin{equation}
    \mathcal D^h_M = \{ \eta^h(v): v \in \{0,1\}^{2M}\}.
\end{equation}
Similarly, we can also define the hole Dirac expansion as the linear combination of elements from $\mathcal D^h_M$. Also, since we restrict ourselves to the case of particle-preserving systems, the basis for the corresponding operator space is given as
\begin{equation}\label{eq: PP_hole_operator_basis}
    \mathcal B^h_M = \left\{ \eta^h(v) \in \mathcal D^h_M : \sum_{k=1}^Mv_{2k-1}=\sum_{k=1}^Mv_{2k}\right\}.
\end{equation}

Within this restriction, the number of annihilation and number of creation operators are always equal, and hence we do not have to define a different weight expression. Incidentally, the proof for weight monotonicity was also independent of the basis. However, we now change the rule for weight truncation. If the initial state has at most $\ell=(M-N)$ holes, we can safely truncate strings with $w(v)>\ell$ since any string with more than $\ell$ creation operators will give $0$ overlap with the initial state. Hence, in the dense filling regime where $\ell$ is small, the truncation scheme within the hole basis is favorable over the canonical basis. The multiplication process is also agnostic to the change of basis due to our choice of bit mapping in the hole basis. Given these observations, we always use the canonical basis for $N\leq M/2$ and the hole basis for $N>M/2$. Thus, within a particle-preserving simulation, the number of terms is bounded by the term scaling given in \eqref{eq:relevant-string-count}.

\section{Expressions}\label{app:expressions}
In this section, we provide the expressions for the behavior of canonical Dirac strings $U^\dagger\eta(v)U$ under several generators. We first describe the effect on single creation and annihilation operators, and then generalize to generic strings $\eta(v)$ by using
\begin{equation}
  U^\dagger\eta(v)U
  =
  \prod_{k=1}^{M}
  \left(U^\dagger f_k^\dagger U\right)^{v_{2k-1}}
  \left(U^\dagger f_k U\right)^{v_{2k}}.
  \label{eq:hopping-full-string}
\end{equation}
After expanding the products, each term is restored to canonical
mode order using the fermionic reordering sign and the local multiplication table.

The general splitting rule is given by \eqref{eq:splitting-rule}, and we recap it here. We have
\begin{equation}
\begin{split}
    U^\dagger_\pm\eta(v)U_\pm
    =\;&\eta(v)
    +\I\frac{\sin(\sqrt{\kappa}\theta)}{\sqrt{\kappa}}
      [G_\pm,\eta(v)]\\
    &+\frac{\cos(\sqrt{\kappa}\theta)-1}{\kappa}
      [G_\pm,[G_\pm,\eta(v)]],
\end{split}
\label{eq:splitting-rule-app}
\end{equation}
where for $\eta(g)\in\mathcal B_M$ we introduced
\begin{align}\label{eq:string-generator-app}
G_+(g)&\equiv\eta(g)+\eta(g)^\dagger \\
G_-(g)&\equiv-\I(\eta(g)-\eta(g)^\dagger),
\end{align}
and $U_\pm=e^{-\I\theta G_\pm(g)}$. In order to obtain the splitting rule from the Hadamard series \eqref{eq:hadamard-series}, we need
\begin{equation}
    \operatorname{ad}_{G_\pm}^3\big(\eta(v)\big)
    =\kappa\,\operatorname{ad}_{G_\pm}\big(\eta(v)\big),
    \qquad \kappa\in\{1,4\}.
    \label{eq:commutator-closure-app}
\end{equation}
Furthermore, the value of $\kappa$ is determined by
\begin{align}
\kappa=1, & \text { if } G_\pm[\eta(v), G_\pm] G_\pm=0, \\
\kappa=4, & \text { if } G_\pm[\eta(v), G_\pm] G_\pm=-[\eta(v), G_\pm].
\end{align}
Below we report the explicit results of the single-mode splitting rules for some of the more common fermionic terms.

\subsection{Hopping generator} 
We consider a hopping term
\begin{equation}
  H_{ij}=f_i^\dagger f_j+f_j^\dagger f_i,
  \qquad
  U_{ij}(\theta)=e^{-\I\theta H_{ij}}.
\end{equation}
The Heisenberg action on the creation operators is
\begin{equation}\label{eq:expansion_series}
  U_{ij}^\dagger f_k^\dagger U_{ij}
  =
  \begin{cases}
    \cos\theta\,f_i^\dagger+\I\sin\theta\,f_j^\dagger,
      & k=i,\\
    \cos\theta\,f_j^\dagger+\I\sin\theta\,f_i^\dagger,
      & k=j,\\
    f_k^\dagger,
      & k\notin\{i,j\}.
  \end{cases}
\end{equation}
The commutators read
\begin{align}
    [H_{ij}, f^\dagger_k] = [f^\dagger_i f_j, f_k^\dagger] + [f^\dagger_j f_i, f_k^\dagger].
\end{align}
We consider first $k \notin \{ i, j\}$. We have
\begin{equation}
    [f^\dagger_i f_j, f_k^\dagger] = 0,
\end{equation}
whereas for $k = j$
\begin{equation}
    [f^\dagger_i f_j, f_j^\dagger] = f_i^\dagger,
\end{equation}
and similarly for $k=i$.
Altogether, 
\begin{equation}
    [H_{ij}, f^\dagger_k] = \begin{cases}
        f^\dagger_j & k = i \\ f^\dagger_i & k = j\\ 0 & k \notin \{i, j\}
    \end{cases}
\end{equation}
which leads to \eqref{eq:expansion_series}. 

\subsection{Density generator}

Consider a density generator
\begin{equation}
  H_i=n_i=f_i^\dagger f_i,
  \qquad
  U_i(\theta)=e^{-\I\theta H_i}.
  \label{eq:density-generator}
\end{equation}
We then find
\begin{equation}
  U_i^\dagger f_k^\dagger U_i
  =
  \begin{cases}
    e^{\I\theta}f_i^\dagger,
      & k=i,\\
    f_k^\dagger,
      & k\neq i.
  \end{cases}
  \label{eq:density-creation}
\end{equation}

\subsection{Density-density generator}
For a density-density generator on two distinct modes $i$ and $j$, let
\begin{equation}
  H_{ij}=n_i n_j,
  \qquad
  U_{ij}(\theta)=e^{-\I\theta H_{ij}}.
  \label{eq:density-density-generator}
\end{equation}
The Heisenberg action on the creation operators is
\begin{equation}
  U_{ij}^{\dagger}f_k^\dagger U_{ij}
  =
  \begin{cases}
    f_i^\dagger+
    \left(e^{\I\theta}-1\right)f_i^\dagger n_j,
      & k=i,\\
    f_j^\dagger+
    \left(e^{\I\theta}-1\right)f_j^\dagger n_i,
      & k=j,\\
    f_k^\dagger,
      & k\notin\{i,j\}.
  \end{cases}
  \label{eq:density-density-creation}
\end{equation}

\subsection{Double-excitation generator}

Consider now a double-excitation process between four pairwise distinct modes $p,q,r,s$. Define
\begin{align}
  A_{pqrs}&=f_p^\dagger f_q^\dagger f_r f_s \\
  H_{pqrs}
  &=A_{pqrs}+A_{pqrs}^\dagger
\end{align}
with corresponding unitary $U_{pqrs}(\theta)=e^{-\I\theta H_{pqrs}}$.
For compactness, define the occupation projectors
\begin{align}
  Q_p
  &=
  n_q(\id-n_r)(\id-n_s)
  +(\id-n_q)n_rn_s,
  \notag\\
  Q_q
  &=
  n_p(\id-n_r)(\id-n_s)
  +(\id-n_p)n_rn_s,
  \notag\\
  Q_r
  &=
  n_s(\id-n_q)(\id-n_p)
  +(\id-n_s)n_qn_p,
  \notag\\
  Q_s
  &=
  n_r(\id-n_q)(\id-n_p)
  +(\id-n_r)n_qn_p.
  \label{eq:four-fermion-projectors}
\end{align}
The nontrivial Heisenberg actions on the creation operators are
\begin{align}
U_{pqrs}^\dagger f_k^\dagger U_{pqrs}
={}&f_k^\dagger
+\I\sin\theta\Bigl(
  \delta_{kp}f_s^\dagger f_r^\dagger f_q
 -\delta_{kq}f_s^\dagger f_r^\dagger f_p
 \Bigr)
\notag\\
&+\I\sin\theta\Bigl(
 -\delta_{kr}f_p^\dagger f_q^\dagger f_s
 +\delta_{ks}f_p^\dagger f_q^\dagger f_r
 \Bigr)
\notag\\
&+(\cos\theta-1)f_k^\dagger
 \Bigl(
   \delta_{kp}Q_p+\delta_{kq}Q_q
 \Bigr) \notag \\
 &+(\cos\theta-1)f_k^\dagger
 \Bigl(
  \delta_{kr}Q_r+\delta_{ks}Q_s
 \Bigr).
\label{eq:four-fermion-creation}
\end{align}

\subsection{Density-assisted hopping generator}

Density-assisted hopping arises from four-fermion terms with one repeated
index. For three distinct modes $i,j,\ell$, define $U_{ij|\ell}(\theta)=e^{-\I\theta H_{ij|\ell}}$ with
\begin{equation}
  H_{ij|\ell}
  =
  n_\ell
  \left(f_i^\dagger f_j+f_j^\dagger f_i\right)
  =n_\ell H_{ij}.
  \label{eq:density-assisted-hopping-generator}
\end{equation}
Because $n_\ell$ is a projector and commutes with $H_{ij}$, the unitary can be recast into the form
\begin{equation}
  U_{ij|\ell}(\theta)
  =
  (\id-n_\ell)+n_\ell e^{-\I\theta H_{ij}}.
  \label{eq:controlled-hopping-unitary}
\end{equation}
Thus, hopping between $i$ and $j$ occurs only when mode $\ell$ is occupied.
We then find
\begin{multline}
U_{ij|\ell}^\dagger f_k^\dagger U_{ij|\ell}
= \\
\begin{cases}
f_i^\dagger
+(\cos\theta-1)f_i^\dagger n_\ell
+\I\sin\theta\,f_j^\dagger n_\ell,
&k=i,\\[2pt]
f_j^\dagger
+(\cos\theta-1)f_j^\dagger n_\ell
+\I\sin\theta\,f_i^\dagger n_\ell,
&k=j,\\[2pt]
f_\ell^\dagger e^{\I\theta H_{ij}},
&k=\ell,\\[2pt]
f_k^\dagger,
&k\notin\{i,j,\ell\}.
\end{cases}
\label{eq:density-assisted-hopping-creation}
\end{multline}
The exponential appearing in the $k=\ell$ case can be written explicitly as
\begin{equation}
  e^{\I\theta H_{ij}}
  =
  \id+\I\sin\theta\,H_{ij}
  +(\cos\theta-1)
  \left(n_i+n_j-2n_in_j\right),
  \label{eq:hopping-exponential-closed}
\end{equation}
where we used
\begin{equation}
  H_{ij}^2=n_i+n_j-2n_in_j.
\end{equation}

\section{Truncation errors}\label{sec:coeff_trunc_proof}
In DP, the Dirac expansion is truncated after a unitary channel is applied. Drawing from existing work on error bounds on the observables in PP~\cite{rudolph2025paulipropagationcomputationalframework}, we show that similar bounds hold for DP as well. Given an initial $N$-particle state $\rho_0$ and an initial observable $O$, consider applying a sequence of unitary channels $\mathcal{C} = \mathcal{C}_L \circ \mathcal{C}_{L-1} \circ \ldots \circ \mathcal{C}_1$. The expectation value for an observable $O$ and an initial state $\rho_0$ is given as 
\begin{align}
    \langle O \rangle &= \Tr(O \mathcal{C}(\rho_0))\\
    &= \Tr(\mathcal{C}^\dagger (O) \rho_0).
\end{align}
In practice, we truncate terms after having applied each channel and denote the new truncated observable after the $j$-th channel application as $O_{L-j}$. The truncated terms are thus given as
\begin{equation}\label{eq:truncated_terms}
    \mathcal{C}_{L-j+1}^\dagger(O_{L-j+1})-O_{L-j}=\sum_{v} d_{j,v} \eta(v)
\end{equation}
where $O_{L}\equiv O$.
It is important to note that coefficient and weight truncation preserve Hermiticity, and thus the truncated Dirac expansion in \eqref{eq:truncated_terms} is Hermitian as well.

As pointed out in the main text, weight truncation for weights above the particle sector does not introduce errors in the expectation value. The error on the expectation value at layer $(L-j)$ due to coefficient truncation only is given as
\begin{equation}\label{eq:DeltaL}
    \Delta_{L-j} = |\mathrm{Tr}_N[(\mathcal{C}_{L-j+1}^\dagger(O_{L-j+1})-O_{L-j})\rho_{L-j}]|
\end{equation}
where $\rho_{L-j} = \mathcal{C}_{L-j}\circ \ldots \circ \mathcal{C}_1(\rho_0)$. Using \eqref{eq:truncated_terms},
\begin{equation}
\begin{split}
    \Delta_{L-j}&=\left|\sum_{v}  \mathrm{Tr}_N(d_{j,v} \eta(v)\rho_{L-j})\right|\\
    &=  \left|\sum_{v}\frac{1}{2}\mathrm{Tr}_N((d_{j,v}\eta(v) + d_{j,v}^* \eta(v)^\dagger )\rho_{L-j})\right|\\
    &\leq \sum_{\substack{v:\\\mathrm{unpaired}\\\mathrm{strings}}} \frac{1}{2} \left(|  \mathrm{Re}(d_{j,v})|+|\mathrm{Im}(d_{j,v})|\right)+\sum_{\substack{v:\\\mathrm{paired}\\\mathrm{strings}}} |d_{j,v}|\label{eq:error_triangle_inequality}
\end{split}
\end{equation}
where we have used the triangle inequality in \eqref{eq:error_triangle_inequality}. We also used 
\begin{align}
    &\mathrm{Tr}_N((\eta(v)+\eta(v)^\dagger)\rho_{L-j})\leq 1 \\
    &\mathrm{Tr}_N(\I(\eta(v)-\eta(v)^\dagger)\rho_{L-j})\leq 1,
\end{align}
since the Hermitian sums for unpaired Dirac strings have eigenvalues $\pm 1$. The argument for fully paired Dirac strings has to be considered separately, because we get $\eta(v)=\eta(v)^\dagger$.

If we apply the Cauchy--Schwarz inequality instead, we get
\begin{equation}
\begin{split}
    \Delta_{L-j}^2 &= \mathrm{Tr}_N((\mathcal{C}_{L-j+1}^\dagger(O_{L-j+1})-O_{L-j})\rho_{L-j})^2 \\ & \leq ||\mathcal{C}_{L-j+1}^\dagger(O_{L-j+1})-O_{L-j}||^2_F \, \mathrm{Tr}_N(\rho_{L-j}^2).
    \end{split}
\end{equation}
As $\rho_{L-j}$ is a pure $N$-particle state, $\mathrm{Tr}_N(\rho_{L-j}^2)=1$. Thus,
\begin{equation}\label{eq:Delta_2norm}
    \Delta_{L-j} \leq ||\mathcal{C}_{L-j+1}^\dagger(O_{L-j+1})-O_{L-j}||_F.
\end{equation}

Using the inner product expression of Dirac expansions, we get
\begin{equation}
    ||\mathcal{C}_{L-j+1}^\dagger(O_{L-j+1})-O_{L-j}||_F = \sqrt{\sum_{u, v} G_{u v}d_{j,u}^*d_{j,v}}.
\end{equation}

The total error $\Delta$ can be bounded by iterating at each layer and using the triangle inequality,
\begin{align}
    \Delta \leq \sum_{l=0}^{L-1}\Delta_{l},
\end{align}
using either \eqref{eq:error_triangle_inequality} or \eqref{eq:Delta_2norm}.
The Cauchy--Schwarz bound \eqref{eq:Delta_2norm} is looser compared to the triangle inequality \eqref{eq:error_triangle_inequality} for computing expectations. However, the Cauchy--Schwarz bounds are tighter for operator overlaps.

\bibliography{biblio}

@article{MagoulasEvangelista2026,
  author        = {Magoulas, Ilias and Evangelista, Francesco A.},
  title         = {Clifford transformations for fermionic quantum systems:
                   From {Pauli} and {Majorana} operators to {Dirac} fermions},
  journal       = {Physical Review A},
  volume        = {113},
  number        = {2},
  pages         = {022455},
  year          = {2026},
  month         = feb,
  publisher     = {American Physical Society},
  doi           = {10.1103/y52r-pd4y},
  eprint        = {2510.23923},
  archiveprefix = {arXiv},
  primaryclass  = {quant-ph},
  url           = {https://doi.org/10.1103/y52r-pd4y}
}

@misc{rycroft_theoretical_physics_2,
  author       = {Slager, R.-J. and Chaudhary, G.},
  title        = {Theoretical Physics II: Topics in Quantum Theory},
  howpublished = {University of Cambridge, Department of Physics},
  year         = {2024},
  note         = {Appendix A, Eq.~(A.2), Hadamard lemma},
  url          = {https://www.tcm.phy.cam.ac.uk/~rjs269/TP2/TP2y2024.pdf}
}

@article{DAnnaNysCarrasquilla2026,
  author        = {D'Anna, Matteo and Nys, Jannes and Carrasquilla, Juan},
  title         = {Majorana string simulation of nonequilibrium dynamics in
                   two-dimensional lattice fermion systems},
  journal       = {Physical Review B},
  volume        = {114},
  number        = {3},
  pages         = {034305},
  year          = {2026},
  month         = jul,
  publisher     = {American Physical Society},
  doi           = {10.1103/f6mn-jc3r},
  eprint        = {2511.02809},
  archiveprefix = {arXiv},
  primaryclass  = {cond-mat.quant-gas},
  url           = {https://doi.org/10.1103/f6mn-jc3r}
}

@misc{MillerEtAl2025MajoranaPropagation,
  author        = {Miller, Aaron and Favre, Joachim and Holmes, Zo{\"e} and
                   Salehi, {\"O}zlem and Chakraborty, Rahul and
                   Nyk{\"a}nen, Anton and Zimbor{\'a}s, Zolt{\'a}n and
                   Glos, Adam and Garc{\'i}a-P{\'e}rez, Guillermo},
  title         = {Simulation of {F}ermionic circuits using
                   {M}ajorana {P}ropagation},
  year          = {2025},
  eprint        = {2503.18939},
  archiveprefix = {arXiv},
  primaryclass  = {quant-ph},
  doi           = {10.48550/arXiv.2503.18939},
  url           = {https://arxiv.org/abs/2503.18939}
}

@article{BegusicChan2025,
  author = {Begu{\v{s}}i{\'c}, Tomislav and Chan, Garnet Kin-Lic},
  title = {Real-Time Operator Evolution in Two and Three
           Dimensions via Sparse {Pauli} Dynamics},
  journal = {PRX Quantum},
  volume = {6},
  pages = {020302},
  year = {2025},
  doi = {10.1103/PRXQuantum.6.020302}
}

@article{EvangelistaMagoulas2025,
  author = {Evangelista, Francesco A. and Magoulas, Ilias},
  title = {Exact closed-form unitary transformations of
           fermionic operators},
  journal = {Phys. Rev. A},
  volume = {111},
  pages = {042825},
  year = {2025},
  doi = {10.1103/PhysRevA.111.042825}
}

@article{DiVincenzoTerhal2005,
  author = {DiVincenzo, David P. and Terhal, Barbara M.},
  title = {Fermionic Linear Optics Revisited},
  journal = {Found. Phys.},
  volume = {35},
  pages = {1967--1984},
  year = {2005},
  doi = {10.1007/s10701-005-8657-0}
}

@article{LiEtAl2020RealTime,
  author = {Li, Xiaosong and Govind, Niranjan and
            Isborn, Christine and DePrince, III, A. Eugene and
            Lopata, Kenneth},
  title = {Real-Time Time-Dependent Electronic Structure Theory},
  journal = {Chemical Reviews},
  volume = {120},
  number = {18},
  pages = {9951--9993},
  year = {2020},
  doi = {10.1021/acs.chemrev.0c00223}
}

@article{YabanaBertsch1996,
  author = {Yabana, K. and Bertsch, G. F.},
  title = {Time-dependent local-density approximation in real time},
  journal = {Physical Review B},
  volume = {54},
  pages = {4484--4487},
  year = {1996},
  doi = {10.1103/PhysRevB.54.4484}
}

@article{XuEtAl2025Hubbard,
  author = {Xu, Muqing and Kendrick, Lev Haldar and Kale, Anant
            and Gang, Youqi and Feng, Chunhan and Zhang, Shiwei
            and Young, Aaron W. and Lebrat, Martin
            and Greiner, Markus},
  title = {A neutral-atom {Hubbard} quantum simulator
           in the cryogenic regime},
  journal = {Nature},
  volume = {642},
  pages = {909--915},
  year = {2025},
  doi = {10.1038/s41586-025-09112-w}
}

@article{BojovicEtAl2026Gates,
  author = {Bojovi{\'c}, Petar and Hilker, Timon and Wang, Si
            and Obermeyer, Johannes and Barendregt, Marnix
            and Tell, Dorothee and Chalopin, Thomas
            and Preiss, Philipp M. and Bloch, Immanuel
            and Franz, Titus},
  title = {High-fidelity collisional quantum gates
           with fermionic atoms},
  journal = {Nature},
  volume = {652},
  pages = {602--608},
  year = {2026},
  doi = {10.1038/s41586-026-10356-3}
}

@article{Rall2019Simulation,
  title = {Simulation of qubit quantum circuits via Pauli propagation},
  author = {Rall, Patrick and Liang, Daniel and Cook, Jeremy and Kretschmer, William},
  journal={Physical Review A},
  volume = {99},
  issue = {6},
  pages = {062337},
  numpages = {10},
  year = {2019},
  month = {Jun},
  publisher = {American Physical Society},
  doi = {10.1103/PhysRevA.99.062337},
  url = {https://link.aps.org/doi/10.1103/PhysRevA.99.062337}
}

@article{
begusic2024Fast,
author = {Tomislav Begušić  and Johnnie Gray  and Garnet Kin-Lic Chan},
title = {Fast and converged classical simulations of evidence for the utility of quantum computing before fault tolerance},
journal = {Science Advances},
volume = {10},
number = {3},
pages = {eadk4321},
year = {2024},
doi = {10.1126/sciadv.adk4321},
URL = {https://www.science.org/doi/abs/10.1126/sciadv.adk4321},
eprint = {https://www.science.org/doi/pdf/10.1126/sciadv.adk4321}}

@article{PolkovnikovEtAl2011,
  author = {Polkovnikov, Anatoli and Sengupta, Krishnendu and
            Silva, Alessandro and Vengalattore, Mukund},
  title = {Colloquium: Nonequilibrium dynamics of closed
           interacting quantum systems},
  journal = {Rev. Mod. Phys.},
  volume = {83},
  pages = {863--883},
  year = {2011},
  doi = {10.1103/RevModPhys.83.863}
}

@article{KirilyukEtAl2010,
  author = {Kirilyuk, Andrei and Kimel, Alexey V. and Rasing, Theo},
  title = {Ultrafast optical manipulation of magnetic order},
  journal = {Rev. Mod. Phys.},
  volume = {82},
  pages = {2731--2784},
  year = {2010},
  doi = {10.1103/RevModPhys.82.2731}
}

@misc{rudolph2025paulipropagationcomputationalframework,
      title={Pauli Propagation: A Computational Framework for Simulating Quantum Systems}, 
      author={Manuel S. Rudolph and Tyson Jones and Yanting Teng and Armando Angrisani and Zoë Holmes},
      year={2025},
      eprint={2505.21606},
      archivePrefix={arXiv},
      primaryClass={quant-ph},
      url={https://arxiv.org/abs/2505.21606},
      doi={10.48550/arXiv.2505.21606}
}

@article{Bach_2019,
  title = {Orthogonalization of fermion $k$-body operators and representability},
  author = {Bach, Volker and Rauch, Robert},
  journal = {Phys. Rev. A},
  volume = {99},
  issue = {4},
  pages = {042109},
  numpages = {10},
  year = {2019},
  month = {Apr},
  publisher = {American Physical Society},
  doi = {10.1103/PhysRevA.99.042109},
  url = {https://link.aps.org/doi/10.1103/PhysRevA.99.042109}
}

@article{Brezinova2024,
  title = {Dynamical quasicondensation in the weakly interacting Fermi-Hubbard model},
  author = {B{\ifmmode \check{r}\else \v{r}\fi{}}ezinov{\'a}, Iva and Stimpfle, Markus and Donsa, Stefan and Rubio, Angel},
  journal = {Phys. Rev. B},
  volume = {109},
  issue = {17},
  pages = {174308},
  numpages = {14},
  year = {2024},
  month = {May},
  publisher = {American Physical Society},
  doi = {10.1103/PhysRevB.109.174308},
  url = {https://link.aps.org/doi/10.1103/PhysRevB.109.174308}
}

@article{kandala2017hardware,
  author  = {Kandala, Abhinav and Mezzacapo, Antonio and Temme, Kristan and Takita, Maika and Brink, Markus and Chow, Jerry M. and Gambetta, Jay M.},
  title   = {Hardware-efficient variational quantum eigensolver for small molecules and quantum magnets},
  journal = {Nature},
  volume  = {549},
  number  = {7671},
  pages   = {242--246},
  year    = {2017},
  doi     = {10.1038/nature23879}
}

@article{hempel2018quantum,
  author  = {Hempel, Cornelius and Maier, Christine and Romero, Jonathan and McClean, Jarrod and Monz, Thomas and Shen, Heng and Jurcevic, Petar and Lanyon, Ben P. and Love, Peter and Babbush, Ryan and Aspuru-Guzik, Al{\'a}n and Blatt, Rainer and Roos, Christian F.},
  title   = {Quantum Chemistry Calculations on a Trapped-Ion Quantum Simulator},
  journal = {Phys. Rev. X},
  volume  = {8},
  number  = {3},
  pages   = {031022},
  year    = {2018},
  doi     = {10.1103/PhysRevX.8.031022}
}

@book{szabo1996modern,
  title={Modern Quantum Chemistry: Introduction to Advanced Electronic Structure Theory},
  author={Szabo, A. and Ostlund, N.S.},
  isbn={9780486691862},
  lccn={88027206},
  series={Dover Books on Chemistry},
  url={https://books.google.co.in/books?id=6mV9gYzEkgIC},
  year={1996},
  publisher={Dover Publications}
}

@book{helgaker2000molecular,
  author    = {Helgaker, Trygve and J{\o}rgensen, Poul and Olsen, Jeppe},
  title     = {Molecular Electronic-Structure Theory},
  publisher = {Wiley},
  address   = {Chichester},
  year      = {2000},
  isbn      = {978-0-471-96755-2},
  doi       = {10.1002/9781119019572}
}

@article{sun2018pyscf,
  author  = {Sun, Qiming and Berkelbach, Timothy C. and Blunt, Nick S. and Booth, George H. and Guo, Sheng and Li, Zhendong and Liu, Junzi and McClain, James D. and Sayfutyarova, Elvira R. and Sharma, Sandeep and Wouters, Sebastian and Chan, Garnet Kin-Lic},
  title   = {{PySCF}: the {Python}-based simulations of chemistry framework},
  journal = {WIREs Comput. Mol. Sci.},
  volume  = {8},
  number  = {1},
  pages   = {e1340},
  year    = {2018},
  doi     = {10.1002/wcms.1340}
}

@article{sun2020recent,
  author  = {Sun, Qiming and Zhang, Xing and Banerjee, Samragni and Bao, Peng and Barbry, Marc and Blunt, Nick S. and Bogdanov, Nikolay A. and Booth, George H. and Chen, Jia and Cui, Zhi-Hao and Eriksen, Janus J. and Gao, Yang and Guo, Sheng and Hermann, Jan and Hermes, Matthew R. and Koh, Kevin and Koval, Peter and Lehtola, Susi and Li, Zhendong and Liu, Junzi and Mardirossian, Narbe and McClain, James D. and Motta, Mario and Mussard, Bastien and Pham, Hung Q. and Pulkin, Artem and Purwanto, Wirawan and Robinson, Paul J. and Ronca, Enrico and Sayfutyarova, Elvira R. and Scheurer, Maximilian and Schurkus, Henry F. and Smith, James E. T. and Sun, Chong and Sun, Shi-Ning and Upadhyay, Shiv and Wagner, Lucas K. and Wang, Xiao and White, Alec and Whitfield, James Daniel and Williamson, Mark J. and Wouters, Sebastian and Yang, Jun and Yu, Jason M. and Zhu, Tianyu and Berkelbach, Timothy C. and Sharma, Sandeep and Sokolov, Alexander Yu. and Chan, Garnet Kin-Lic},
  title   = {Recent developments in the {PySCF} program package},
  journal = {J. Chem. Phys.},
  volume  = {153},
  number  = {2},
  pages   = {024109},
  year    = {2020},
  doi     = {10.1063/5.0006074}
}

@article{hehre1969self,
  author  = {Hehre, W. J. and Stewart, R. F. and Pople, J. A.},
  title   = {Self-Consistent Molecular-Orbital Methods. {I}. Use of {Gaussian} Expansions of {Slater}-Type Atomic Orbitals},
  journal = {J. Chem. Phys.},
  volume  = {51},
  number  = {6},
  pages   = {2657--2664},
  year    = {1969},
  doi     = {10.1063/1.1672392}
}

@article{hehre1970self,
  author  = {Hehre, W. J. and Ditchfield, R. and Stewart, R. F. and Pople, J. A.},
  title   = {Self-Consistent Molecular Orbital Methods. {IV}. Use of {Gaussian} Expansions of {Slater}-Type Orbitals. Extension to Second-Row Molecules},
  journal = {J. Chem. Phys.},
  volume  = {52},
  number  = {5},
  pages   = {2769--2773},
  year    = {1970},
  doi     = {10.1063/1.1673374}
}

@article{ditchfield1971self,
  author  = {Ditchfield, R. and Hehre, W. J. and Pople, J. A.},
  title   = {Self-Consistent Molecular-Orbital Methods. {IX}. An Extended {Gaussian}-Type Basis for Molecular-Orbital Studies of Organic Molecules},
  journal = {J. Chem. Phys.},
  volume  = {54},
  number  = {2},
  pages   = {724--728},
  year    = {1971},
  doi     = {10.1063/1.1674902}
}

@article{hehre1972self,
  author  = {Hehre, W. J. and Ditchfield, R. and Pople, J. A.},
  title   = {Self-Consistent Molecular Orbital Methods. {XII}. Further Extensions of {Gaussian}-Type Basis Sets for Use in Molecular Orbital Studies of Organic Molecules},
  journal = {J. Chem. Phys.},
  volume  = {56},
  number  = {5},
  pages   = {2257--2261},
  year    = {1972},
  doi     = {10.1063/1.1677527}
}

@article{dill1975self,
  author  = {Dill, James D. and Pople, John A.},
  title   = {Self-consistent molecular orbital methods. {XV}. Extended {Gaussian}-type basis sets for lithium, beryllium, and boron},
  journal = {J. Chem. Phys.},
  volume  = {62},
  number  = {7},
  pages   = {2921--2923},
  year    = {1975},
  doi     = {10.1063/1.430801}
}

@article{hariharan1973influence,
  author  = {Hariharan, P. C. and Pople, J. A.},
  title   = {The influence of polarization functions on molecular orbital hydrogenation energies},
  journal = {Theor. Chim. Acta},
  volume  = {28},
  number  = {3},
  pages   = {213--222},
  year    = {1973},
  doi     = {10.1007/BF00533485}
}

@article{dunning1989gaussian,
  author  = {Dunning, Jr., Thom H.},
  title   = {{Gaussian} basis sets for use in correlated molecular calculations. {I}. The atoms boron through neon and hydrogen},
  journal = {J. Chem. Phys.},
  volume  = {90},
  number  = {2},
  pages   = {1007--1023},
  year    = {1989},
  doi     = {10.1063/1.456153}
}

@article{prascher2011gaussian,
  author  = {Prascher, Brian P. and Woon, David E. and Peterson, Kirk A. and Dunning, Jr., Thom H. and Wilson, Angela K.},
  title   = {{Gaussian} basis sets for use in correlated molecular calculations. {VII}. Valence, core-valence, and scalar relativistic basis sets for {Li}, {Be}, {Na}, and {Mg}},
  journal = {Theor. Chem. Acc.},
  volume  = {128},
  number  = {1},
  pages   = {69--82},
  year    = {2011},
  doi     = {10.1007/s00214-010-0764-0}
}

@article{cohen1965electric,
  author  = {Cohen, Howard D. and Roothaan, C. C. J.},
  title   = {Electric Dipole Polarizability of Atoms by the {Hartree--Fock} Method. {I}. Theory for Closed-Shell Systems},
  journal = {J. Chem. Phys.},
  volume  = {43},
  number  = {10},
  pages   = {S34--S39},
  year    = {1965},
  doi     = {10.1063/1.1701512}
}

@article{krause2005timedependent,
  author  = {Krause, Pascal and Klamroth, Tillmann and Saalfrank, Peter},
  title   = {Time-dependent configuration-interaction calculations of laser-pulse-driven many-electron dynamics: Controlled dipole switching in lithium cyanide},
  journal = {J. Chem. Phys.},
  volume  = {123},
  number  = {7},
  pages   = {074105},
  year    = {2005},
  doi     = {10.1063/1.1999636}
}

@article{li2020realtime,
  author  = {Li, Xiaosong and Govind, Niranjan and Isborn, Christine and DePrince, III, A. Eugene and Lopata, Kenneth},
  title   = {Real-Time Time-Dependent Electronic Structure Theory},
  journal = {Chem. Rev.},
  volume  = {120},
  number  = {18},
  pages   = {9951--9993},
  year    = {2020},
  doi     = {10.1021/acs.chemrev.0c00223}
}

@article{cederbaum1999ultrafast,
  author  = {Cederbaum, L. S. and Zobeley, J.},
  title   = {Ultrafast charge migration by electron correlation},
  journal = {Chem. Phys. Lett.},
  volume  = {307},
  number  = {3--4},
  pages   = {205--210},
  year    = {1999},
  doi     = {10.1016/S0009-2614(99)00508-4}
}

@article{motta2021low,
  author        = {Motta, Mario and Ye, Erika and McClean, Jarrod R. and Li, Zhendong and Minnich, Austin J. and Babbush, Ryan and Chan, Garnet Kin-Lic},
  title         = {Low rank representations for quantum simulation of electronic structure},
  journal       = {npj Quantum Inf.},
  volume        = {7},
  pages         = {83},
  year          = {2021},
  doi           = {10.1038/s41534-021-00416-z}
}

@article{Keldysh:1964ud,
    author = "Keldysh, L. V.",
    title = "{Diagram Technique for Nonequilibrium Processes}",
    doi = "10.1142/9789811279461_0007",
    journal = "Sov. Phys. JETP",
    volume = "20",
    pages = "1018--1026",
    year = "1965"
}

@article{runge1984density,
  author  = {Runge, Erich and Gross, E. K. U.},
  title   = {Density-Functional Theory for Time-Dependent Systems},
  journal = {Phys. Rev. Lett.},
  volume  = {52},
  number  = {12},
  pages   = {997--1000},
  year    = {1984},
  doi     = {10.1103/PhysRevLett.52.997}
}

@article{abrams1997simulation,
  author  = {Abrams, Daniel S. and Lloyd, Seth},
  title   = {Simulation of Many-Body {Fermi} Systems on a Universal Quantum Computer},
  journal = {Phys. Rev. Lett.},
  volume  = {79},
  number  = {13},
  pages   = {2586--2589},
  year    = {1997},
  doi     = {10.1103/PhysRevLett.79.2586}
}

@article{eckstein2009thermalization,
  author  = {Eckstein, Martin and Kollar, Marcus and Werner, Philipp},
  title   = {Thermalization after an Interaction Quench in the {Hubbard} Model},
  journal = {Phys. Rev. Lett.},
  volume  = {103},
  number  = {5},
  pages   = {056403},
  year    = {2009},
  doi     = {10.1103/PhysRevLett.103.056403}
}

@article{beaurepaire1996ultrafast,
  author  = {Beaurepaire, E. and Merle, J.-C. and Daunois, A. and Bigot, J.-Y.},
  title   = {Ultrafast Spin Dynamics in Ferromagnetic Nickel},
  journal = {Phys. Rev. Lett.},
  volume  = {76},
  number  = {22},
  pages   = {4250--4253},
  year    = {1996},
  doi     = {10.1103/PhysRevLett.76.4250}
}

@article{aoki2014nonequilibrium,
  author  = {Aoki, Hideo and Tsuji, Naoto and Eckstein, Martin and Kollar, Marcus and Oka, Takashi and Werner, Philipp},
  title   = {Nonequilibrium dynamical mean-field theory and its applications},
  journal = {Rev. Mod. Phys.},
  volume  = {86},
  number  = {2},
  pages   = {779--837},
  year    = {2014},
  doi     = {10.1103/RevModPhys.86.779}
}

@article{bauer2020quantum,
  author  = {Bauer, Bela and Bravyi, Sergey and Motta, Mario and Chan, Garnet Kin-Lic},
  title   = {Quantum Algorithms for Quantum Chemistry and Quantum Materials Science},
  journal = {Chem. Rev.},
  volume  = {120},
  number  = {22},
  pages   = {12685--12717},
  year    = {2020},
  doi     = {10.1021/acs.chemrev.9b00829}
}

@article{bauer2023highenergy,
  author  = {Bauer, Christian W. and Davoudi, Zohreh and Balantekin, A. Baha and Bhattacharya, Tanmoy and Carena, Marcela and de Jong, Wibe A. and Draper, Patrick and El-Khadra, Aida and Gemelke, Nate and Hanada, Masanori and Kharzeev, Dmitri and Lamm, Henry and Li, Ying-Ying and Liu, Junyu and Lukin, Mikhail D. and Meurice, Yannick and Monroe, Christopher and Nachman, Benjamin and Pagano, Guido and Preskill, John and Rinaldi, Enrico and Roggero, Alessandro and Santiago, David I. and Savage, Martin J. and Siddiqi, Irfan and Siopsis, George and Van Zanten, David and Wiebe, Nathan and Yamauchi, Yukari and Yeter-Aydeniz, K{\"u}bra and Zorzetti, Silvia},
  title   = {Quantum Simulation for High-Energy Physics},
  journal = {PRX Quantum},
  volume  = {4},
  pages   = {027001},
  year    = {2023},
  doi     = {10.1103/PRXQuantum.4.027001}
}

@article{Bourgund_2025,
   title={Formation of individual stripes in a mixed-dimensional cold-atom Fermi–Hubbard system},
   volume={637},
   ISSN={1476-4687},
   url={http://dx.doi.org/10.1038/s41586-024-08270-7},
   DOI={10.1038/s41586-024-08270-7},
   number={8044},
   journal={Nature},
   publisher={Springer Science and Business Media LLC},
   author={Bourgund, Dominik and Chalopin, Thomas and Bojović, Petar and Schlömer, Henning and Wang, Si and Franz, Titus and Hirthe, Sarah and Bohrdt, Annabelle and Grusdt, Fabian and Bloch, Immanuel and Hilker, Timon A.},
   year={2025},
   month=Jan, pages={57–62} }

@article{Hirthe_2023,
   title={Magnetically mediated hole pairing in fermionic ladders of ultracold atoms},
   volume={613},
   ISSN={1476-4687},
   url={http://dx.doi.org/10.1038/s41586-022-05437-y},
   DOI={10.1038/s41586-022-05437-y},
   number={7944},
   journal={Nature},
   publisher={Springer Science and Business Media LLC},
   author={Hirthe, Sarah and Chalopin, Thomas and Bourgund, Dominik and Bojović, Petar and Bohrdt, Annabelle and Demler, Eugene and Grusdt, Fabian and Bloch, Immanuel and Hilker, Timon A.},
   year={2023},
   month=Jan, pages={463–467} }

@article{moeckel2010crossover,
  author  = {Moeckel, Michael and Kehrein, Stefan},
  title   = {Crossover from adiabatic to sudden interaction quenches in the {Hubbard} model: prethermalization and non-equilibrium dynamics},
  journal = {New J. Phys.},
  volume  = {12},
  number  = {5},
  pages   = {055016},
  year    = {2010},
  doi     = {10.1088/1367-2630/12/5/055016}
}

@article{baiardi2021electron,
  author  = {Baiardi, Alberto},
  title   = {Electron Dynamics with the Time-Dependent Density Matrix Renormalization Group},
  journal = {J. Chem. Theory Comput.},
  volume  = {17},
  number  = {6},
  pages   = {3320--3334},
  year    = {2021},
  doi     = {10.1021/acs.jctc.0c01048}
}

@article{nascimento2016linear,
  author  = {Nascimento, Daniel R. and DePrince, III, A. Eugene},
  title   = {Linear Absorption Spectra from Explicitly Time-Dependent Equation-of-Motion Coupled-Cluster Theory},
  journal = {J. Chem. Theory Comput.},
  volume  = {12},
  number  = {12},
  pages   = {5834--5840},
  year    = {2016},
  doi     = {10.1021/acs.jctc.6b00796}
}

@article{kivlichan2018quantum,
  author  = {Kivlichan, Ian D. and McClean, Jarrod and Wiebe, Nathan and Gidney, Craig and Aspuru-Guzik, Al{\'a}n and Chan, Garnet Kin-Lic and Babbush, Ryan},
  title   = {Quantum Simulation of Electronic Structure with Linear Depth and Connectivity},
  journal = {Phys. Rev. Lett.},
  volume  = {120},
  number  = {11},
  pages   = {110501},
  year    = {2018},
  doi     = {10.1103/PhysRevLett.120.110501}
}

@article{peng2018dynamics,
  author  = {Peng, Wei-Tao and Fales, B. Scott and Levine, Benjamin G.},
  title   = {Simulating Electron Dynamics of Complex Molecules with Time-Dependent Complete Active Space Configuration Interaction},
  journal = {J. Chem. Theory Comput.},
  volume  = {14},
  number  = {8},
  pages   = {4129--4138},
  year    = {2018},
  doi     = {10.1021/acs.jctc.8b00381}
}

@article{rocca2024symmetry,
  author  = {Rocca, Dario and Cortes, Cristian L. and Gonthier, J{\'e}r{\^o}me F. and Ollitrault, Pauline J. and Parrish, Robert M. and Anselmetti, Gian-Luca and Degroote, Matthias and Moll, Nikolaj and Santagati, Raffaele and Streif, Michael},
  title   = {Reducing the Runtime of Fault-Tolerant Quantum Simulations in Chemistry through Symmetry-Compressed Double Factorization},
  journal = {J. Chem. Theory Comput.},
  volume  = {20},
  number  = {11},
  pages   = {4639--4653},
  year    = {2024},
  doi     = {10.1021/acs.jctc.4c00352}
}

@article{rigol2008thermalization,
  author  = {Rigol, Marcos and Dunjko, Vanja and Olshanii, Maxim},
  title   = {Thermalization and its mechanism for generic isolated quantum systems},
  journal = {Nature},
  volume  = {452},
  pages   = {854--858},
  year    = {2008},
  doi     = {10.1038/nature06838}
}

@article{bruner2017attosecond,
  author  = {Bruner, Adam and Hernandez, Samuel and Mauger, Fran{\c{c}}ois and Abanador, Paul M. and LaMaster, Daniel J. and Gaarde, Mette B. and Schafer, Kenneth J. and Lopata, Kenneth},
  title   = {Attosecond Charge Migration with {TDDFT}: Accurate Dynamics from a Well-Defined Initial State},
  journal = {J. Phys. Chem. Lett.},
  volume  = {8},
  number  = {17},
  pages   = {3991--3996},
  year    = {2017},
  doi     = {10.1021/acs.jpclett.7b01652}
}

@article{lee2021hypercontraction,
  author  = {Lee, Joonho and Berry, Dominic W. and Gidney, Craig and Huggins, William J. and McClean, Jarrod R. and Wiebe, Nathan and Babbush, Ryan},
  title   = {Even More Efficient Quantum Computations of Chemistry Through Tensor Hypercontraction},
  journal = {PRX Quantum},
  volume  = {2},
  number  = {3},
  pages   = {030305},
  year    = {2021},
  doi     = {10.1103/PRXQuantum.2.030305}
}

@article{fontana_classical_2025,
  title = {Classical simulations of noisy variational quantum circuits},
  volume = {11},
  issn = {2056-6387},
  url = {https://doi.org/10.1038/s41534-024-00955-1},
  doi = {10.1038/s41534-024-00955-1},
  number = {1},
  journal = {npj Quantum Information},
  author = {Fontana, Enrico and Rudolph, Manuel S. and Duncan, Ross and Rungger, Ivan and Cîrstoiu, Cristina},
  month = may,
  year = {2025},
  pages = {84},
}

@article{Dirac:1927dy,
    author = "Dirac, Paul A. M.",
    title = "{Quantum theory of emission and absorption of radiation}",
    doi = "10.1098/rspa.1927.0039",
    journal = "Proc. Roy. Soc. Lond. A",
    volume = "114",
    pages = "243",
    year = "1927"
}

@misc{facelli2026fastconvergencemajoranapropagation,
      title={Fast convergence of Majorana Propagation for weakly interacting fermions}, 
      author={Giorgio Facelli and Hamza Fawzi and Omar Fawzi},
      year={2026},
      eprint={2601.05226},
      archivePrefix={arXiv},
      primaryClass={quant-ph},
      url={https://arxiv.org/abs/2601.05226}, 
}

@misc{zhao2026efficientclassicalsimulationweakly,
      title={Efficient Classical Simulation of Weakly Interacting Fermion Dynamics}, 
      author={Chu Zhao and Iman Marvian and Yu Tong},
      year={2026},
      eprint={2608.19448},
      archivePrefix={arXiv},
      primaryClass={quant-ph},
      url={https://arxiv.org/abs/2608.19448}, 
}

@online{MPjl,
  author = {Matteo D'Anna and Manuel S. Rudolph},
  title = {\texttt{MajoranaPropagation.jl}},
  year = 2026,
  url = {https://github.com/SparqleSim/MajoranaPropagation.jl}
}

@article{Hohenstein2012Tensor,
    author = {Hohenstein, Edward G. and Parrish, Robert M. and Martínez, Todd J.},
    title = {Tensor hypercontraction density fitting. I. Quartic scaling second- and third-order Møller-Plesset perturbation theory},
    journal = {The Journal of Chemical Physics},
    volume = {137},
    number = {4},
    pages = {044103},
    year = {2012},
    month = {07},
    issn = {0021-9606},
    doi = {10.1063/1.4732310},
    url = {https://doi.org/10.1063/1.4732310},
}

@article{Parrish2012Tensor,
    author = {Parrish, Robert M. and Hohenstein, Edward G. and Martínez, Todd J. and Sherrill, C. David},
    title = {Tensor hypercontraction. II. Least-squares renormalization},
    journal = {The Journal of Chemical Physics},
    volume = {137},
    number = {22},
    pages = {224106},
    year = {2012},
    month = {12},
    issn = {0021-9606},
    doi = {10.1063/1.4768233},
    url = {https://doi.org/10.1063/1.4768233},
}

@article{Matsuzawa2020Jastrow,
    author = {Matsuzawa, Yuta and Kurashige, Yuki},
    title = {Jastrow-type Decomposition in Quantum Chemistry for
Low-Depth Quantum Circuits},
    journal = {Journal of Chemical Theory and Computation},
    volume = {16},
    number = {2},
    pages = {944-952},
    year = {2020},
    month = {02},
    issn = {1549-9618},
    doi = {10.1021/acs.jctc.9b00963},
    url = {https://doi.org/10.1021/acs.jctc.9b00963},
}

@article{Cohn2021Quantum,
  title = {Quantum Filter Diagonalization with Compressed Double-Factorized Hamiltonians},
  author = {Cohn, Jeffrey and Motta, Mario and Parrish, Robert M.},
  journal = {PRX Quantum},
  volume = {2},
  issue = {4},
  pages = {040352},
  numpages = {19},
  year = {2021},
  month = {Dec},
  publisher = {American Physical Society},
  doi = {10.1103/PRXQuantum.2.040352},
  url = {https://link.aps.org/doi/10.1103/PRXQuantum.2.040352}
}

@article{Oumarou2024acceleratingquantum,
  doi = {10.22331/q-2024-06-13-1371},
  url = {https://doi.org/10.22331/q-2024-06-13-1371},
  title = {Accelerating {Q}uantum {C}omputations of {C}hemistry {T}hrough {R}egularized {C}ompressed {D}ouble {F}actorization},
  author = {Oumarou, Oumarou and Scheurer, Maximilian and Parrish, Robert M. and Hohenstein, Edward G. and Gogolin, Christian},
  journal = {{Quantum}},
  issn = {2521-327X},
  publisher = {{Verein zur F{\"{o}}rderung des Open Access Publizierens in den Quantenwissenschaften}},
  volume = {8},
  pages = {1371},
  month = jun,
  year = {2024}
}

@misc{chakraborty2026scalablequantumcircuitgeneration,
      title={Scalable quantum circuit generation for iterative ground state approximation using Majorana Propagation}, 
      author={Rahul Chakraborty and Aaron Miller and Anton Nykänen and Özlem Salehi and Fabio Tarocco and Fabijan Pavošević and Pi. A. B. Haase and Martina Stella and Adam Glos},
      year={2026},
      eprint={2603.23444},
      archivePrefix={arXiv},
      primaryClass={quant-ph},
      url={https://arxiv.org/abs/2603.23444}, 
}

@article{Shrikhande2026,
  title = {Pauli Propagation Enables Fast Classical Simulation of Strongly Correlated Quantum Systems},
  author = {Shrikhande, Chinmay and Bachhar, Arnab and Rodr{\'{\i}}guez-Jim{\'e}nez, Jos{\'e} Aar{\'o}n and Valeev, Edward F. and Mayhall, Nicholas J.},
  journal = {PRX Quantum},
  volume = {7},
  issue = {4},
  pages = {043022},
  numpages = {25},
  year = {2026},
  month = {Nov},
  publisher = {American Physical Society},
  doi = {10.1103/kq3v-whcq},
  url = {https://link.aps.org/doi/10.1103/kq3v-whcq}
}

@article{Lin2026,
   title={Utility-Scale Quantum State Preparation: Classical Training using Pauli Path Simulation},
   volume={10},
   ISSN={2521-327X},
   url={http://dx.doi.org/10.22331/q-2026-03-09-2014},
   DOI={10.22331/q-2026-03-09-2014},
   journal={Quantum},
   publisher={Verein zur Forderung des Open Access Publizierens in den Quantenwissenschaften},
   author={Lin, Cheng-Ju and Gharibyan, Hrant and Su, Vincent P.},
   year={2026},
   month=Mar, pages={2014} }

@misc{santona2026hardwareefficientgroundstatepreparationusing,
      title={Hardware-Efficient Ground-State Preparation using Variational Imaginary-Time Majorana Evolution}, 
      author={Federico Santona and Manuel G. Algaba and Aeishah Ameera Anuar and Anna M. Wernbacher and Prachi Sharma and Fedor Šimkovic IV},
      year={2026},
      eprint={2610.01954},
      archivePrefix={arXiv},
      primaryClass={quant-ph},
      url={https://arxiv.org/abs/2610.01954}, 
}

@misc{rudolph2023classicalsurrogatesimulationquantum,
      title={Classical surrogate simulation of quantum systems with LOWESA}, 
      author={Manuel S. Rudolph and Enrico Fontana and Zoë Holmes and Lukasz Cincio},
      year={2023},
      eprint={2308.09109},
      archivePrefix={arXiv},
      primaryClass={quant-ph},
      url={https://arxiv.org/abs/2308.09109}, 
}

\end{document}